\documentclass[aps,twocolumn,nofootinbib]{revtex4}
\usepackage{bm,amssymb,epsf}
\newcommand{\kB}{k_{\rm B}}
\newcommand{\Cco}{{\cal C}}

\newcommand{\Kop}{{\cal K}}

\newcommand{\Kopi}{{\cal K}^{-1}}
\newcommand{\Lop}{{\cal L}}
\newcommand{\Lopz}{{\cal L}^{(0)}}
\newcommand{\Mco}{{\cal M}}

\newcommand{\Rop}{{\cal R}}
\newcommand{\Ropz}{{\cal R}^{(0)}}

\newcommand{\ave}[1]{\left\langle#1\right\rangle}
\newcommand{\aves}[1]{\langle#1\rangle}

\newcommand{\cancor}[2]{\left\langle#1;#2\right\rangle}
\newcommand{\ccancor}[2]{\langle\!\langle#1;#2\rangle\!\rangle}
\newcommand{\Qcommu}[2]{[#1,#2]}
\newcommand{\QcommuB}[2]{\Big[#1,#2\Big]}

\begin{document}
\bibliographystyle{apsrev}

\title{Dynamic Coarse-Graining Approach to Quantum Field Theory}

\author{Hans Christian \"Ottinger}
\email[]{hco@mat.ethz.ch}
\homepage[]{http://www.polyphys.mat.ethz.ch/}
\affiliation{ETH Z\"urich, Department of Materials, Polymer Physics, HCI H 543,
CH-8093 Z\"urich, Switzerland}

\date{\today}

\begin{abstract}
We build quantum field theory on the thermodynamic master equation for dissipative quantum systems. The vacuum is represented by a thermodynamic equilibrium state in the low-temperature limit. All regularization is consistently provided by a friction mechanism; with decreasing friction parameter, only degrees of freedom on shorter and shorter length scales are damped out of a quantum field theory. No divergent integrals need to be manipulated. Renormalization occurs as a tool to refine perturbation expansions, not to remove divergences. Relativistic covariance is recovered in the final results. We illustrate the proposed thermodynamic approach to quantum fields for the $\varphi^4$ theory by calculating the propagator and the $\beta$ function, and we offer some suggestions on its application to gauge theories.
\end{abstract}



\maketitle

\section{Introduction}
Renormalization plays a key role in quantum field theory. The renormalization group is intrinsically related to the successive elimination of degrees of freedom and should hence be expected to result in coarse-grained equations. Quantum field theory hence belongs into the world of multiscale modeling and coarse graining of dynamic systems. The field idealization implies the need to bridge a wide range of length scales and, in relativistic quantum field theory, clearly also in time scales. We hence need a proper framework to implement dynamic renormalization-group theory.

The proper setting for multiscale modeling and coarse graining is statistical nonequilibrium thermodynamics. Whereas, in the context of quantum field theory, one usually does not think of irreversible processes, the creation and annihilation of particle-antiparticle pairs at inaccessibly small length and time scales, for example, certainly is beyond our control and should hence be regarded and treated as an irreversible process. We hence propose to make use of the powerful machinery of nonequilibrium thermodynamics instead of inventing a set of sophisticated rules to eliminate various kinds of infinities from an approach designed for reversible systems.

As a result of the thermodynamic approach, we avoid the possible concern that renormalization ``is simply a way to sweep the difficulties of the divergences of [quantum] electrodynamics under the rug,'' as expressed by Feynman in a catchy metaphorical statement in his Nobel lecture (1965). Or, in the words of the insistent critic Dirac \cite{Diracrem}, ``the quantum mechanics that most physicists are using nowadays [in quantum field theory] is just a set of working rules, and not a complete dynamical theory at all.'' Whereas Dirac felt the need for a different type of Hamiltonian, we here suggest to address the intrinsic irreversibility associated with the field idealization and renormalization in an appropriate manner. Maybe such a thermodynamic approach introducing irreversibility into quantum field theory could eventually provide ``some really drastic changes'' in the equations, as demanded by Dirac (see pp.~36-37 of \cite{Diraced}).

We start by introducing some relevant background information: The quantum master equation required to formulate the dissipative dynamics of quantum systems and the renormalization-group approach employed to refine the results of perturbation theory. We then develop the thermodynamic approach for a scalar quantum field with quartic interactions. It is shown how the interactions can be handled by a straightforward, thermodynamically consistent perturbation theory that takes detailed balance into account; first and second order results for the propagator and a four-point correlation are calculated and discussed in great detail. The steps required to generalize the dissipative approach from scalar to more complicated fields, such as Yang-Mills gauge fields, are sketched in a further section. We conclude with a brief summary and some further remarks on the coarse-graining approach to quantum field theory. Useful details are compiled in a number of appendices. Throughout this paper, we use units with $\hbar=c=1$ where $\hbar$ is the reduced Planck constant and $c$ is the speed of light.

\section{Quantum master equation}
In this section, we compile the equations required to describe the mixed reversible-irreversible evolution of quantum systems so that we can later formulate a dissipative smoothing mechanism for quantum field theories. After introducing the general framework, we focus on the equations obtained by linearization around an equilibrium state, on detailed balance, and the resulting guidance for the development of perturbation theory.

\subsection{Thermodynamic master equation}
A standard approach to quantum dissipation is based on linear quantum master equations for density matrices \cite{BreuerPetru,Weiss}. The form of the most general linear quantum master equation that is guaranteed to preserve the trace and positive-semidefiniteness of any initial density matrix has been determined by Lindblad \cite{Lindblad76}. However, master equations of the Lindblad form are not applicable to arbitrarily low temperatures. As we wish to construct quantum field theories in the limit of zero temperature and vanishing dissipation rate, we need to pay attention to the thermodynamic consistency of the quantum master equation to render it applicable for low temperatures and weak friction \cite{Grabert06}.

Motivated by a failure of the ``quantum regression hypothesis,'' Grabert revisited the standard projection-operator derivation of quantum master equations \cite{Grabert} with a relevant density matrix of the exponential form, where the deviation from the Hamiltonian in the exponent can be interpreted as the thermodynamic force operator conjugate to the density matrix. In the Markovian limit, the resulting equation for the evolution of the density matrix or statistical operator $\rho$ is of the nonlinear form (see Eq.~(5.22) of \cite{Grabert82}),
\begin{eqnarray}
    \frac{d\rho}{dt} &=& i \Qcommu{\rho}{H}
    - \sum_{j,k} \gamma_{jk} \Qcommu{Q_j}{\Kop_\rho \Qcommu{Q_k}{H}} \nonumber \\
    &-& \kB T_{\rm e} \sum_{j,k} \gamma_{jk} \, \Qcommu{Q_j}{\Qcommu{Q_k}{\rho}} .
\label{Grabertme}
\end{eqnarray}
The reversible first term on the right-hand side of Eq.~(\ref{Grabertme}) has the well-known representation in terms of the commutator and the Hamiltonian $H$. The additional irreversible term is formulated in terms of a suitable matrix of damping coefficients $\gamma_{jk}$ describing the strength of the dissipation and the observables $Q_j$ describing the interaction between the quantum subsystem and its environment which, in the case of quantum field theory, is a bath representing the unresolvable or eliminated small-scale features. As usual, $\kB$ is Boltzmann's constant and $T_{\rm e}$ is the temperature. Nonlinearity arises from the super-operator
\begin{equation}\label{condensmat}
    \Kop_\rho A = \int_0^1 \rho^u A \, \rho^{1-u} \, du ,
\end{equation}
which adds a factor of $\rho$ to the observable $A$ with a proper treatment of ordering problems. Note that the temperature $T_{\rm e}$ is the only parameter characterizing the state of the environment, which can hence be regarded as a heat bath. Equation (\ref{Grabertme}) may be addressed as a thermodynamic master equation because it has been derived with a relevant density matrix characterized in terms of a thermodynamic force operator and because, as a consequence, it is consistent with the fluctuation-dissipation theorem.

The quantum master equation (\ref{Grabertme}) and its generalization for the coupling to arbitrary classical nonequilibrium systems as environments has recently been obtained in \cite{hco199}. Starting point is the geometric formulation of classical nonequilibrium thermodynamics in terms of Poisson and dissipative brackets \cite{hco99,hco100,hcobet}. The quantum generalization is obtained by Dirac's method of classical analogy \cite{Dirac}: Poisson brackets are replaced by commutators, dissipative brackets are replaced by canonical correlations of commutators. The appealing properties of the nonlinear quantum master equation (\ref{Grabertme}) have been discussed in \cite{hco201}. As the purely phenomenological and formal projection operator approaches to quantum dissipation lead to the same nonlinear thermodynamic master equation applicable in the low-temperature regime, we have a save starting point for our dissipative approach to quantum field theory.

\subsection{Linearized master equation}
In order to clarify the nonlinear nature of the thermodynamic quantum master equation (\ref{Grabertme}) and to linearize it around its equilibrium solution
\begin{equation}\label{eqsolqme}
    \rho_{\rm eq} = \frac{e^{- \beta H}}{{\rm tr} \, e^{- \beta H}} ,
\end{equation}
at inverse temperature $\beta = (\kB T_{\rm e})^{-1}$, we introduce the thermodynamic driving force operator,
\begin{equation}\label{thermdriveforce}
    \mu = \kB T_{\rm e} \, ( \ln \rho - \ln \rho_{\rm eq} ) .
\end{equation}
We can then write
\begin{equation}\label{tdriveid}
    \Kop_\rho \Qcommu{Q_k}{H} = \Kop_\rho \Qcommu{Q_k}{\mu}
    - \kB T_{\rm e} \Qcommu{\Kop_\rho  Q_k}{\ln \rho} .
\end{equation}
With this simple consequence of the definition of the thermodynamic driving force operator and the useful identity
\begin{equation}\label{lnLemma}
     \Qcommu{A}{\rho} = \Qcommu{\Kop_\rho A}{\ln\rho}
     = \Kop_\rho \Qcommu{A}{\ln\rho} ,
\end{equation}
which follows from looking at arbitrary matrix elements formed with the eigenstates of the density matrix and performing the elementary integration over $u$ in Eq.~(\ref{condensmat}), we can rewrite the quantum master equation (\ref{Grabertme}) in the more compact form
\begin{equation}\label{Grabertmemu}
    \frac{d\rho}{dt} = i \Qcommu{\rho}{H}
    - \sum_{j,k} \gamma_{jk} \Qcommu{Q_j}{\Kop_\rho \Qcommu{Q_k}{\mu}} .
\end{equation}
In view of the logarithmic form of $\mu$ given in Eq.~(\ref{thermdriveforce}), this equation is clearly nonlinear in $\rho$. It moreover justifies the interpretation of $\mu$ as a thermodynamic driving force operator.

In linearizing around equilibrium, the driving force $\mu$ is small so that $\Kop_\rho$ in Eq.~(\ref{Grabertmemu}) can be evaluated with the equilibrium density matrix; in that case, we use the symbol $\Kop$. From the linearized relationship between the deviation from the equilibrium density matrix and the thermodynamic driving force operator,
\begin{equation}\label{thermdriveforcexp}
    \rho =\exp \{ \ln \rho_{\rm eq} + \beta \mu \}
    \approx \rho_{\rm eq} + \Kop \beta \mu ,
\end{equation}
we finally obtain the linearized thermodynamic master equation in the form
\begin{equation}\label{Grabertmelin}
    \frac{d\rho}{dt} = i \Qcommu{\rho}{H}
    - \kB T_{\rm e} \sum_{j,k} \gamma_{jk} \Qcommu{Q_j}{\Kop \Qcommu{Q_k}{\Kopi \rho}} .
\end{equation}
Note that the master equation (\ref{Grabertmelin}) is not of the Lindblad form \cite{Lindblad76}. For large deviations from equilibrium, positive-semidefiniteness of the density matrix is only guaranteed by the nonlinear terms. For small deviations from equilibrium, we can safely use Eq.~(\ref{Grabertmelin}).

\subsection{Adjointness properties}
For the general development of the theory of linearized thermodynamic quantum master equations, it is convenient to look at the time evolution of the observable $\hat{\rho} = \Kopi\rho$. In other words, $\hat{\rho}$ is the relative deviation of the density matrix $\rho$ from its equilibrium form, $\rho_{\rm eq}$. By operating with $\Kopi$ on the linearized quantum master equation (\ref{Grabertmelin}), the evolution equation for the observable $\hat{\rho}$ is given by
\begin{equation}\label{Grabertmelinrel}
    \frac{d\hat{\rho}}{dt} = \Lop\hat{\rho} = i \Qcommu{\hat{\rho}}{H}
    - \kB T_{\rm e} \sum_{j,k} \gamma_{jk} \Kopi \Qcommu{Q_j}{\Kop \Qcommu{Q_k}{\hat{\rho}}} .
\end{equation}
Of course, $\hat{\rho}=1$ is found to be the steady state solution of this equation.

The average of any observable $A$ can be obtained as
\begin{equation}\label{avdefforms}
    {\rm tr}(A \rho) = {\rm tr}(A \Kop \hat{\rho}) = \cancor{A}{\hat{\rho}} ,
\end{equation}
where the second identity is the definition of the equilibrium canonical correlation (see Eq.~(4.1.12) of \cite{KuboetalII} or Eq.~(3) of \cite{hco199}). The time evolution of an average can be obtained from the master equation (\ref{Grabertmelin}) for the evolution of $\rho$ or from Eq.~(\ref{Grabertmelinrel}) for $\hat{\rho}$. According to the property
\begin{equation}\label{adjointLprop}
    \cancor{A}{\Lop B} = \cancor{\bar{\Lop} A}{B} ,
\end{equation}
where the adjoint operator $\bar{\Lop}$ follows from an elementary calculation,
\begin{equation}\label{adjointLdef}
    \bar{\Lop} A = - i \Qcommu{A}{H}
    - \kB T_{\rm e} \sum_{j,k} \gamma_{kj} \Kopi \Qcommu{Q_j}{\Kop \Qcommu{Q_k}{A}} ,
\end{equation}
there exists a third possibility to obtain the evolution of an average: one can use the time-dependent observable $A$ obtained from $d A / dt = \bar{\Lop} A$ and a constant initial density matrix $\rho$ or $\hat{\rho}$ in Eq.~(\ref{avdefforms}). The reversible contribution to the evolution equation for $A$ corresponds to the usual Heisenberg equation for observables. The operator $\bar{\Lop}$ is the adjoint of $\Lop$ in canonical correlations and the adjoint of the operator appearing on the right-hand side of the quantum master equation (\ref{Grabertmelin}) under the plain trace operation. For symmetric matrices $\gamma_{jk}$, the only difference between the operators $\Lop$ and $\bar{\Lop}$ is in the sign of the reversible term.

The adjointness property (\ref{adjointLprop}) establishes a relationship or consistency between time evolution and canonical equilibrium correlations. It may hence be regarded as a detailed-balance condition that ensures the proper symmetry of two-time canonical correlations.

\subsection{Detailed balance}\label{secmulcor}
The usual construction of multi-time correlations relies on the possibility of introducing a Heisenberg picture, that is, on the use of time-dependent operators averaged with a time-independent density matrix \cite{BreuerPetru}. For a nonlinear master equation governing the evolution of the density matrix in the Schr\"odinger picture, the passage to a Heisenberg picture is no longer possible. We are hence faced with a serious problem when we wish to study multi-time correlations. Even for the linearized master equation, one needs to be careful with employing the Heisenberg-like picture to define multi-time correlations. To discuss this problem in more detail, we start from the identity
\begin{equation}\label{lnLemmatr}
     {\rm tr}( \Qcommu{A}{B} \rho ) =
     {\rm tr}\Big\{ (\Kop_\rho A) \, \Qcommu{B}{\ln\rho} \Big\} ,
\end{equation}
which follows from Eq.~(\ref{lnLemma}). When applied to the equilibrium density matrix (\ref{eqsolqme}), we obtain the relation
\begin{equation}\label{comcangenid}
    \ave{\Qcommu{A}{B}} = i \beta \cancor{A}{\Lop_{\rm rev} B}
    = i \beta \cancor{\bar{\Lop}_{\rm rev} A}{B} ,
\end{equation}
between equilibrium averages and canonical correlations, where $\Lop_{\rm rev}$ and $\bar{\Lop}_{\rm rev} = - \Lop_{\rm rev}$ are the reversible contributions to the super-operators $\Lop$ and $\bar{\Lop}$ defined in Eqs.~(\ref{Grabertmelinrel}) and (\ref{adjointLdef}). After replacing $A$ by $e^{\bar{\Lop} t} A$, we obtain the rigorous identity
\begin{equation}\label{preFDR}
    \ave{\Qcommu{e^{\bar{\Lop} t} A}{B}} =
    i \beta \cancor{\bar{\Lop}_{\rm rev} e^{\bar{\Lop} t} A}{B} .
\end{equation}
For purely reversible dynamics, $\bar{\Lop} = \bar{\Lop}_{\rm rev}$, we finally obtain
\begin{equation}\label{FDR}
    \ave{\Qcommu{e^{\bar{\Lop} t} A}{B}} =
    i \beta \, \frac{d}{d t} \cancor{e^{\bar{\Lop} t} A}{B} ,
\end{equation}
which is known as a fluctuation-dissipation relation between a response function and a correlation function (see, for example, \cite{Grabert82} or Eq.~(4.2.18) of \cite{KuboetalII}). As it has been derived for from the fundamental Hamiltonian equation of motion, where $A(t) = e^{\bar{\Lop} t} A$ is the time-evolving operator of the Heisenberg picture, the fluctuation-dissipation relation (\ref{FDR}) should be respected by all coarse-grained evolution equations, for example, the quantum master equation.

For $\bar{\Lop} \neq \bar{\Lop}_{\rm rev}$, it is not allowed to define various two-time correlations naively by inserting $e^{\bar{\Lop} t} A$ because Eq.~(\ref{preFDR}) would then imply a violation of the fluctuation-dissipation relation. This subtlety is known as the failure of the quantum regression hypothesis \cite{Grabert82,GraberTalkner83,Talkner86,hco201}. It is related to the fact that an adjointness property like in Eq.~(\ref{adjointLprop}) does not generally exist for the average of commutators,
\begin{eqnarray}
    && \hspace{-1.2cm} \ave{\Qcommu{\bar{\Lop} A}{B}} - \ave{\Qcommu{A}{\Lop B}} \, =
    \nonumber \\
    && i \beta \Big( \cancor{\bar{\Lop}_{\rm rev} \bar{\Lop} A}{B}
    - \cancor{A}{\Lop_{\rm rev} \Lop B} \Big)  \, =
    \nonumber \\
    && i \beta \cancor{A}{( \Lop \Lop_{\rm rev} - \Lop_{\rm rev} \Lop ) B} ,
\label{shuffelid2}
\end{eqnarray}
which follows from Eqs.~(\ref{comcangenid}) and (\ref{adjointLprop}). Adjointness, or detailed balance, in commutator averages can be guaranteed only for purely reversible dynamics.

\subsection{Perturbation theory}
In the following, we are interested in frequency-dependent correlations rather than in the time-dependent correlation function (\ref{FDR}). We hence introduce the super-operator
\begin{equation}\label{Ropbardef}
    \bar{\Rop}(\omega) = \int_0^\infty e^{(\bar{\Lop} -i \omega) t} \, dt .
\end{equation}
If necessary for convergence, the frequency $\omega$ can have a small negative imaginary part. For nontrivially interacting systems, we would like to calculate such correlation functions by means of perturbation theory. We hence assume that the total Hamiltonian can be written as a sum, $H = H^{(0)} + H^{(1)}$, where for the super-operator $\bar{\Lop}^{(0)}$ obtained for $H^{(0)}$ instead of $H$ in the definition (\ref{adjointLdef}) (note that also $\Kop$ involves $H$ through the canonical equilibrium density matrix), but with the same coupling operators $Q_k$, the frequency-dependent super-operator characterizing the free evolution,
\begin{equation}\label{Ropbarzdef}
    \bar{\Rop}^{(0)}(\omega) = \int_0^\infty e^{(\bar{\Lop}^{(0)} -i \omega) t} \, dt ,
\end{equation}
can be evaluated explicitly. The operator $\Ropz(\omega)$ is defined in the same way in terms of $\Lopz$. These definitions imply the useful identities
\begin{equation}\label{L0R0ident}
    \Lopz + [\Ropz(\omega)]^{-1} =
    \bar{\Lop}^{(0)} + [\bar{\Rop}^{(0)}(\omega)]^{-1} = i \omega .
\end{equation}

We now consider the straightforward formal second-order perturbation series for the frequency-dependent canonical correlation,
\begin{eqnarray}
    \cancor{\bar{\Rop}(\omega) A}{B} &=&
    \cancor{\bar{\Rop}^{(0)}(\omega) A}{B} \nonumber\\
    &+& \cancor{\bar{\Rop}^{(0)}(\omega) \bar{\Lop}'
    \bar{\Rop}^{(0)}(\omega) A}{B}
    \nonumber\\
    &+& \cancor{\bar{\Rop}^{(0)}(\omega) \bar{\Lop}'
    \bar{\Rop}^{(0)}(\omega) \bar{\Lop}' \bar{\Rop}^{(0)}(\omega) A}{B}
    \nonumber\\
    &+& \ldots
\label{perturbationbasic}
\end{eqnarray}
where we have introduced the following pair of super-operators describing interaction effects,
\begin{equation}\label{Lopprimedef}
    \Lop' = \Lop - \Lopz , \qquad \bar{\Lop}' = \bar{\Lop} - \bar{\Lop}^{(0)} .
\end{equation}
Whereas the adjointness property (\ref{adjointLprop}) implies that the exact result is equal to $\cancor{A}{\Rop(\omega) B}$, such detailed balance properties do not exist for the individual terms of the perturbation expansion. The construction of a proper perturbation theory is even more subtle than the implementation of the fluctuation-dissipation relation because the perturbation expansion involves multi-time correlations. In order to arrive at a more symmetric formulation, we need to find a suitable adjointness property involving $\Lop'$ and $\bar{\Lop}'$. From Eqs.~(\ref{adjointLprop}) and (\ref{L0R0ident}), we obtain the adjointness property
\begin{eqnarray}
    \cancor{\bar{\Lop}' \bar{\Rop}^{(0)}(\omega) A}{\Ropz(\omega) B}
    + \cancor{\bar{\Rop}^{(0)}(\omega) A}{B} = &&
    \nonumber \\
    \cancor{\bar{\Rop}^{(0)}(\omega) A}{\Lop' \Ropz(\omega) B}
    + \cancor{A}{\Ropz(\omega) B} . &&
\label{adjointLprimeprop}
\end{eqnarray}
Moving $\Lop'$ to the other side is linked with moving $\Ropz(\omega)$ to the other side in a lower-order term. The structure of terms of successive orders in perturbation theory becomes coupled. After moving the first factor $\bar{\Rop}^{(0)}(\omega)$ in the highest-order term of Eq.~(\ref{perturbationbasic}), the adjointness property (\ref{adjointLprimeprop}) allows us to rewrite the perturbation expansion in the more symmetric form
\begin{eqnarray}
    \cancor{\bar{\Rop}(\omega) A}{B} &=&
    \cancor{\bar{\Rop}^{(0)}(\omega) A}{B} \nonumber\\
    &+& \cancor{\bar{\Lop}'
    \bar{\Rop}^{(0)}(\omega) A}{\Ropz(\omega) B}
    \nonumber\\
    &+& \cancor{\bar{\Rop}^{(0)}(\omega) \bar{\Lop}'
    \bar{\Rop}^{(0)}(\omega) A}{\Lop' \Ropz(\omega) B}
    \nonumber\\
    &+& \ldots
\label{perturbationbasicsym}
\end{eqnarray}
The occurrence of super-operators acting on $A$ and $B$ is now balanced.

In practical calculations, averages of commutators are more convenient than canonical correlations. By analogy with Eq.~(\ref{perturbationbasicsym}) we hence introduce the following correlation function guided by symmetry in time,
\begin{eqnarray}
    \Cco_{AB}(\omega) &=&
    \ave{\Qcommu{\bar{\Rop}^{(0)}(\omega) A}{B}} \nonumber\\
    &+& \ave{\Qcommu{\bar{\Lop}'
    \bar{\Rop}^{(0)}(\omega) A}{\Ropz(\omega) B}}
    \nonumber\\
    &+& \ave{\Qcommu{\bar{\Rop}^{(0)}(\omega) \bar{\Lop}'
    \bar{\Rop}^{(0)}(\omega) A}{\Lop' \Ropz(\omega) B}}
    \nonumber\\
    &+& \ldots
\label{Cdefperturbation}
\end{eqnarray}
For Hamiltonian dynamics, with $\Lop = \Lop_{\rm rev}$, this perturbation expansion can be obtained in exactly the same way as Eq.~(\ref{perturbationbasicsym}) because Eq.~(\ref{shuffelid2}) provides the required adjointness property. For general master equations, with $\Lop \neq \Lop_{\rm rev}$, we proceed as for the fluctuation-dissipation relation and postulate that the balanced perturbation expansion (\ref{Cdefperturbation}) still must be applicable. The validity of this postulate can be verified in the applications of Eq.~(\ref{Cdefperturbation}) in Secs.~\ref{subsecprop} and \ref{subsecfourpt}.

A more rigorous approach to time-reversal and detailed balance properties in perturbation theory will be of particular importance if one is interested in multi-time correlations, such as Wilson loops. We need a generalization of the discussion of time-reversal symmetry and detailed balance for the fluctuation-dissipation relation in Secs.~4 and 6 of \cite{Grabert82}. For our purposes, however, the balanced second-order perturbation expansion (\ref{Cdefperturbation}) turns out to be fully satisfactory.

The averages in Eq.~(\ref{Cdefperturbation}) are still for the interacting theory. In the following, we need the first-order expansion (see, for example, Eq.~(4.1.10) of \cite{KuboetalII}, or Eq.~(5.1.9) of \cite{Grabert})
\begin{equation}\label{averageperturb}
    \ave{A} = \ave{A}^{(0)} - \beta \ccancor{A}{H^{(1)}}^{(0)} + \ldots ,
\end{equation}
where the notation $\ccancor{\cdot}{\cdot}^{(0)}$ implies that, in the evaluation of the free canonical correlation by means of Wick's theorem, at least one contraction between $A$ and $H^{(1)}$ needs to be present. In other words, the contributions from a product of averages is suppressed so that we might call $\ccancor{\cdot}{\cdot}^{(0)}$ a canonical covariance.

\section{Renormalization group}
Renormalization is sometimes perceived as a tricky toolbox to remove annoying divergences from quantum field theory. The present section emphasizes that the renormalization group should rather be considered as a profound tool to refine perturbation expansions which would actually be useless without renormalization. We first try to develop some intuition in the context of polymer physics where no divergences are present. We then present a few equations that provide practical recipes for refining perturbation expansions and allow us to calculate the famous $\beta$ function for the running coupling constants of renormalization-group theory.

\subsection{Intuitive example}
Plain perturbation theory is clearly not appropriate for problems involving a large number of interactions. However, if a problem exhibits self-similarity on different length scales, perturbation theory can be refined to obtain useful results by successively accounting for infinitely many interactions. This refinement may be considered as a kind of generalized exponentiation procedure guided by a renormalization-group analysis.

Nice illuminating examples of refined perturbation expansions can be found in the theory of linear polymer molecules. The beauty of polymer physics actually stems from the self-similarity of polymers \cite{deGennes}. If we model polymer molecules in dilute solution as linear chains of beads connected by springs, hydrodynamic interactions between the beads arise because each bead perturbs the solvent flow around it and, after propagation of the perturbation, it affects the motion of the other beads. The bead friction coefficient determines the strength of such hydrodynamic-interaction effects.

The beads of such mechanical polymer models, however, are fictitious objects consisting of many monomers. If one uses larger beads, hydrodynamic interactions between the smaller beads inside a larger bead have to be incorporated into the effective friction coefficient of the larger beads. This consideration offers the possibility of incorporating more and more interactions by passing to successively larger beads. Small increments in bead size implying only few interactions inside the larger beads can be handled by perturbation theory and renormalization-group theory allows us to accumulate a very large number of interactions via many small steps. A number of static and dynamic properties of polymers in solution have been computed by these ideas \cite{Oono85,Freed,desCloizeauxJannink,hco38,hco39,hco185}. The transformation behavior between effective interaction strengths on different length scales contains important information, in particular, about the critical exponents associated with self-similarity and the limiting value of the effective coupling strength on large scales. With the large scale model, one can finally perform a perturbative calculation of any quantity of interest. A comparison with the scaling expressions for these quantities following from self-similarity allows us to refine the results of plain perturbation theory even further (see, for example, the behavior of various material functions at high shear rates discussed in \cite{hco43}).

Note that divergences are not an issue in the above discussion of hydrodynamic interactions in dilute polymer solutions. They would only arise if, to establish contact with field theory, we considered the limit of an infinitely large number of infinitesimally small beads. In the next section, we show how the intuitive ideas of this section can be translated into tractable equations.

\subsection{Basic equations}
Let us introduce a small length scale $\ell$, which could be a bead size, a lattice spacing, an inverse cutoff for momentum or frequency, or the characteristic length scale of dissipative smoothing. If some model of interest contains a dimensional coupling constant $\lambda$, say the strength of some interaction, the proper choice of which typically depends on $\ell$, we first introduce the dimensionless coupling constant
\begin{equation}\label{lambdadimless}
    \tilde{\lambda}(\ell) = \ell^\epsilon \lambda(\ell) ,
\end{equation}
with suitable exponent $\epsilon$ obtained from dimensional analysis. The intuitive ideas described in the preceding subsection are implemented by constructing a \emph{perturbation theory of the rate of change} of $\tilde{\lambda}(\ell)$ with $\ell$ rather than for $\tilde{\lambda}(\ell)$ itself. We hence assume that there actually exists a perturbation expansion of the function describing the rate of change of the dimensionless coupling constant,
\begin{equation}\label{betadef}
    \beta(\tilde{\lambda}) = - \ell \, \frac{d\tilde{\lambda}}{d\ell} ,
\end{equation}
that is, for the standard $\beta$ function for the running coupling constant of quantum field theory. We always display the $\beta$ function with its argument to avoid confusion with the inverse temperature. If the free theory remains free on all length scales, $\beta(0)=0$, the most general second-order expansion of the $\beta$ function is given by
\begin{equation}\label{beta2nd}
    \beta(\tilde{\lambda}) = - \alpha \tilde{\lambda} \left( 1 - \frac{\tilde{\lambda}}{\lambda^*} \right) ,
\end{equation}
where the parameters $\alpha$ and $\lambda^*$ remain to be determined. According to Eq.~(\ref{betadef}), the second-order perturbation theory (\ref{beta2nd}) for $\beta$ implies the following nontrivial dependence of the running coupling constant on the length scale,
\begin{equation}\label{lambdaform}
    \lambda(\ell) = \frac{\lambda^* \, \ell^{\alpha-\epsilon}}{\ell^\alpha+c} .
\end{equation}
Note that the combination
\begin{equation}\label{lambdaformc}
    \frac{\lambda^*}{c} = \ell^{\alpha-\epsilon} \lambda(\ell)
    \left[ 1 - \frac{\ell^\epsilon \lambda(\ell)}{\lambda^*} \right]^{-1}
\end{equation}
is a constant. Any invariant property $P$ must be some function of this constant and, as a polynomial in $\lambda^*/c$ and eventually in $\lambda(\ell)$, must possess a second-order perturbation expansion of the form
\begin{equation}\label{perturbationform}
    P_0 + P_1 \, \ell^{\alpha-\epsilon} \, \lambda
    \left( 1 + \frac{\ell^\epsilon \lambda}{\lambda^*} \right)
    + P_2 \, \ell^{2(\alpha-\epsilon)} \, \lambda^2 .
\end{equation}
We will later find the identity $\alpha = \epsilon$ so that the perturbation expansion (\ref{perturbationform}) takes an even simpler form. Then, Eq.~(\ref{lambdaform}) shows that $\lambda(0)$ corresponds to $\lambda^*/c$, and $\lambda(\ell) = \ell^{-\epsilon} \lambda^*$ for large $\ell$.

Of course, these arguments can be generalized to construct the most general form of higher-order perturbation expansions of observable quantities. Generalizing Eq.~(\ref{lambdaformc}), a polynomial expansion of the $\beta$ function leads to a nonpolynomial form of the constant,
\begin{equation}\label{lambdahatdef}
    \hat{\lambda} = C \, \ell^{-\epsilon} \exp \left\{ - \epsilon \int^{\tilde{\lambda}}
    \frac{d\tilde{\lambda}'}{\beta(\tilde{\lambda}')} \right\} ,
\end{equation}
where the integration constant $C$ is to be chosen such that $\hat{\lambda} = \lambda$ for small $\ell$. The integral of the rational function $1/\beta(\tilde{\lambda}')$ can actually be performed in closed form, but only an expansion in $\tilde{\lambda}$ is required. Any invariant quantity $P$ must be a function of $\hat{\lambda}$ and hence possesses a perturbation expansion of the form
\begin{equation}\label{genpertexp1}
    P = P_0 + P_1 \hat{\lambda} + P_2 \hat{\lambda}^2 + P_3 \hat{\lambda}^3 \ldots ,
\end{equation}
where the expanded form of Eq.~(\ref{lambdahatdef}) has to be inserted,
\begin{equation}\label{genpertexp2}
    \hat{\lambda} = \lambda + B_2 \ell^\epsilon \lambda^2
    + B_3 \ell^{2\epsilon} \lambda^3 \ldots .
\end{equation}
The combined expansion is of the form
\begin{eqnarray}
    P &=& P_0 + P_1 \lambda + ( P_2 + P_1 B_2 \ell^\epsilon ) \lambda^2 \nonumber\\
    &+& ( P_3 + 2 P_2 B_2 \ell^\epsilon + P_1 B_3 \ell^{2\epsilon} ) \lambda^3 \ldots .
\label{genpertexp3}
\end{eqnarray}
Note that this expansion contains the various corrections from the small length scale $\ell$. The coefficients $P_j$ can easily be read off from the terms of order $\ell^0$ of such an expansion. The coefficients $B_j$ characterizing the $\beta$ function can then be read off in various places; in the language of Feynman diagrams, this possibility corresponds to the occurrence of the same subdiagrams in infinitely many, increasingly complicated diagrams. The consistency of the results and their independence of the particular quantity $P$ expresses the renormalizability of the theory.

Note that the dependence of $\lambda$ on $\ell$ is obtained from the analysis of the perturbative prediction of large scale properties. The parameters $(\lambda(\ell_1), \ell_1)$ and $(\lambda(\ell_2), \ell_2)$ with $\ell_2 > \ell_1$ imply the same large scale properties (on length scales large compared to $\ell_2$), but the model with parameters $(\lambda(\ell_2), \ell_2)$ typically cannot accommodate all properties of the model with parameters $(\lambda(\ell_1), \ell_1)$ on length scales of order $\ell_2$. In other words, the underlying model serves as a minimal model for a universality class without being close to some fixed point model. If we want to refine our perturbation theory for some observable $P$, however, we only have the option to translate the coupling constant $\lambda$ from small scales to some physical length scale, say an inverse mass $m$, and to use the corresponding translated $\lambda$ as $\hat{\lambda}$ in the expansion (\ref{genpertexp1}). An even simpler refinement is obtained if we assume that, at the length scales of interest, $\tilde{\lambda}$ has reached its fixed point value $\lambda^*$ with $\beta(\lambda^*) = 0$, which implies the choice $\hat{\lambda} = m^\epsilon \lambda^*$ in the expansion (\ref{genpertexp1}), or $\lambda = m^\epsilon \lambda^*$ and $\ell = 0$ in the expansion (\ref{genpertexp3}).

Keeping the large scale physics invariant when letting the length scale $\ell$ of the underlying model go to zero implies that the underlying model has to approach a critical point with diverging correlation length in units of $\ell$. This well-known relationship between field theory and critical phenomena, including the calculation of critical exponents, has been discussed extensively, for example, in the nice classical review article \cite{BrezinetalDombGreen}.

\section{$\varphi^4$ theory}
With the tools of quantum dissipation at hand, we can now formulate quantum field theory with small-scale smoothing by a dynamic friction mechanism. Our approach may be considered as a generalization of the canonical quantization procedure to include dissipation. We present the free field theory and construct the propagator and a four-point correlation of the interacting theory in second-order perturbation theory. In a final step, we refine perturbation expansions by means of renormalization-group theory and construct the $\beta$ function.

\subsection{Quantization procedure}
Quantization procedures are traditionally based on a canonical Hamiltonian formulation of the evolution equations for the underlying classical systems. In the canonical approach to quantum mechanics, rooted in Dirac's pioneering work, the canonical Poisson brackets of classical mechanics are replaced by the commutators of quantum mechanics \cite{Diracorig}. This procedure has been adapted to quantum field theory (see, for example, Secs.~11.2 and 11.3 of \cite{BjorkenDrell} or Sec.~I.8 of \cite{Zee}). Even in the path-integral approach to quantum field theory, the justification of the proper action needs to be supported by the canonical approach (see, for example, the introduction to Sec.~9 of \cite{WeinbergQFT1}). Understanding its Hamiltonian structure is hence crucial for quantizing a system.

In classical mechanics, on has the choice between the Lagrangian and Hamiltonian formulations; both formulations can also be used for classical field theories (see, for example, Chaps.~2, 8, and 12 of \cite{Goldstein}). The Lagrangian approach is based directly on the variational principle for the action which is obtained as the time integral of the Lagrangian. In contrast, the equivalent Hamiltonian approach needs two structural elements, the Hamiltonian and a Poisson bracket required to turn the gradient of the Hamiltonian into the vector describing time evolution. The canonical Poisson structure, which in the non-degenerate case is also know as symplectic structure, is the key to formulating the proper commutators in the quantization procedure. It is important to note that the Hamiltonian and the Poisson structure are two separate elements. In particular, once one has identified a fundamental Poisson structure, one can easily change the Hamiltonian. To understand the equivalence of the Lagrangian and Hamiltonian formulations of classical field theory, one should realize that the existence of a non-degenerate Poisson structure is crucial to establish an underlying variational principle \cite{Santilli}.

If one changes the system, one needs to change the Lagrangian and hence one needs to begin from scratch in the Lagrangian approach. In the Hamiltonian approach, one needs to change the Hamiltonian but one may keep the Poisson structure, where the latter is the key to quantization. Of course, one might ask whether we actually want to change the system. This indeed happens when we consider the noninteracting and interacting systems in parallel, say for constructing a perturbation theory. The logical separation of canonical Poisson structures and Hamiltonians is also very useful when one wishes to quantize systems involving nontrivial constraints. Indeed, the proper handling of the constraints resulting from gauge invariance is a major obstacle to quantizing the Yang-Mills and gravitational fields. Note that the Lagrangian and Hamiltonian approaches are closely related to path-integral and canonical quantization, respectively.

As nonequilibrium thermodynamics is built on a Hamiltonian formulation of reversible dynamics, the quantization of dissipative systems is obtained as a generalization of the canonical quantization procedure. As an additional step, one needs to formulate the proper form of the frictional coupling to a heat bath in the quantum master equation.

\subsection{Fields and Hamiltonian}
We consider a scalar quantum field $\varphi(\bm{x})$ and its canonical conjugate $\pi(\bm{x})$ in $d$ space dimensions. Throughout this paper, we denote the dimensions of space and space-time by $d$ and $D=d+1$, respectively. The bosonic field operators $\varphi(\bm{x})$ at all positions $\bm{x}$ commute among each other, and so do the conjugate operators $\pi(\bm{x})$. The only nontrivial commutation relations are of the canonical form
\begin{equation}\label{canoncommutator}
    \Qcommu{\varphi(\bm{x})}{\pi(\bm{x}')} = i \, \delta(\bm{x}-\bm{x}') .
\end{equation}
We assume that the total Hamiltonian of our system is given by
\begin{equation}\label{Hamiltonianresp}
    H = \int \left\{ \frac{1}{2} \left[ \pi^2 + (\nabla \varphi)^2 + m^2 \varphi^2 \right]
    + \frac{1}{24} \lambda \, \varphi^4 \right\} d^dx ,
\end{equation}
where the quadratic contribution in $\pi(\bm{x})$ and $\varphi(\bm{x})$ describes a free massive scalar field with mass parameter $m$ and the quartic contribution is the interaction term of the $\varphi^4$ theory. The parameter $\lambda$, often referred to as a coupling constant, describes the strength of the interaction. A dimensional analysis of Eq.~(\ref{Hamiltonianresp}) shows that $\varphi^2 m^{1-d}$ and $\lambda m^{d-3}$ are dimensionless.

The self-adjoint position-dependent canonical field variables can be expressed in Fourier representations of the form (see, for example, Sec.~I.8 of \cite{Zee} or Sec.~12.1 of \cite{BjorkenDrell})
\begin{equation}\label{phiexpression}
    \varphi(\bm{x}) = \frac{1}{\sqrt{2(2\pi)^d}} \int \frac{d^dk}{\sqrt{\omega_k}}
    \left( a^\dag_{\bm{k}} + a_{-\bm{k}} \right) e^{- i \bm{k} \cdot \bm{x}} ,
\end{equation}
and
\begin{equation}\label{piexpression}
    \pi(\bm{x}) = \frac{i}{\sqrt{2(2\pi)^d}} \int d^dk \sqrt{\omega_k}
    \left( a^\dag_{\bm{k}} - a_{-\bm{k}} \right) e^{- i \bm{k} \cdot \bm{x}} .
\end{equation}
The only requirement for the function $\omega_k$ occurring in this representation is that it assumes equal values for $\bm{k}$ and $-\bm{k}$. With these Fourier representations, we have introduced the adjoint operators $a^\dag_{\bm{k}}$ and $a_{\bm{k}}$ creating and annihilating field quanta of momentum $\bm{k} \in \mathbb{R}^d$, respectively, as primary variables. All creation operators commute among each other, and so do the annihilation operators. The only nontrivial commutation relations for the boson creation and annihilation operators are
\begin{equation}\label{acommutator}
    \Qcommu{a_{\bm{k}}}{a^\dag_{\bm{k}'}} = \delta(\bm{k}-\bm{k}') .
\end{equation}

We assume that the collection of all states created by multiple application of all the operators $a_{\bm{k}}^\dag$ for all $\bm{k} \in \mathbb{R}^d$ on a ground state (which is annihilated by any $a_{\bm{k}}$) is complete. The full Hilbert space factorizes into spaces obtained by repeated application of $a_{\bm{k}}^\dag$ for each mode $\bm{k}$ (see, for example, Secs.~1 and 2 of \cite{FetterWalecka} or Secs.~12.1 and 12.2 of \cite{BjorkenDrell} for more details on the construction of such Fock spaces). The field quantization based on Eq.~(\ref{acommutator}) for creation and annihilation operators is an equivalent alternative to the canonical quantization procedure based on Eq.~(\ref{canoncommutator}).

We next write the total Hamiltonian as the sum $H = H^{(0)} + H^{(1)}$ and express the quadratic free Hamiltonian $H^{(0)}$ and the quartic interaction term $H^{(1)}$ in terms of creation and annihilation operators. Neglecting an irrelevant constant contribution, the free Hamiltonian can be expressed in the simple form
\begin{equation}\label{H0def}
    H^{(0)} = \int \omega_k \, a^\dag_{\bm{k}} a_{\bm{k}} \, d^dk ,
\end{equation}
provided that the momentum-dependent frequencies $\omega_k$ are given by the relativistic dispersion relation
\begin{equation}\label{dispersion}
    \omega_k^2 = \bm{k}^2 + m^2 .
\end{equation}
Once more neglecting constant terms, the interaction term of the $\varphi^4$ theory is given by
\begin{eqnarray}
    H^{(1)} &=& \frac{\lambda}{96} \frac{1}{(2\pi)^d}
    \int \prod_{j=1}^4 \frac{d^dk_j}{\sqrt{\tilde{\omega}_{k_j}}} \,
    \delta(\bm{k}_1+\bm{k}_2+\bm{k}_3+\bm{k}_4) \quad \nonumber \\
    && \Big( a_{-\bm{k}_1} a_{-\bm{k}_2} a_{-\bm{k}_3} a_{-\bm{k}_4}
    + 4 a^\dag_{\bm{k}_1} a_{-\bm{k}_2} a_{-\bm{k}_3} a_{-\bm{k}_4}
    \nonumber \\
    && + 6 a^\dag_{\bm{k}_1} a^\dag_{\bm{k}_2} a_{-\bm{k}_3} a_{-\bm{k}_4} \nonumber \\
    &&
    + 4 a^\dag_{\bm{k}_1} a^\dag_{\bm{k}_2} a^\dag_{\bm{k}_3} a_{-\bm{k}_4}
    + a^\dag_{\bm{k}_1} a^\dag_{\bm{k}_2} a^\dag_{\bm{k}_3} a^\dag_{\bm{k}_4}
    \Big) \nonumber \\
    &+& \lambda z \int \frac{d^dk}{2 \tilde{\omega}_k} \,
    \left( a_{\bm{k}} a_{-\bm{k}} + 2 a^\dag_{\bm{k}} a_{\bm{k}} +
    a^\dag_{\bm{k}} a^\dag_{-\bm{k}} \right) ,
\label{H1defNO}
\end{eqnarray}
where we have chosen to use the normal ordered form of $H^{(1)}$, that is, all creation operators are moved to the left, all annihilation operators are moved to the right. Normal ordering will be very convenient for the subsequent calculations. By inserting the Fourier transform (\ref{phiexpression}) into the Hamiltonian (\ref{Hamiltonianresp}), we would actually obtain $\tilde{\omega}_k = \omega_k$ and $z = z_{\rm no}$ with
\begin{equation}\label{znocontrib}
    z_{\rm no} = \frac{1}{4} \,
    \frac{1}{(2\pi)^d} \int \frac{d^dq}{2 \tilde{\omega}_q} =
    \frac{1}{4} \, m^{d-1} I_1 ,
\end{equation}
where the integral $I_1$ and its Lorentz invariance are discussed in Appendix \ref{app_integrals} [see Eq.~(\ref{IntegralI1})].

We have introduced the generalization from $\omega_k$ to $\tilde{\omega}_k$ with the idea to modify the frequency at very large $|\bm{k}|$ to guarantee the convergence of all integrals occurring in intermediate steps of the calculation and to allow us a direct comparison with more traditional regularization procedures. The details of this regularization at large wave vectors are irrelevant to the present approach. Note that the passage from $\omega_k$ to $\tilde{\omega}_k$ does not change the Hamiltonian structure (only the Hamiltonian), but it destroys the Lorentz invariance of the system.

It is very convenient to allow for values of $z$ that are different from the value occurring naturally from the normal-ordering procedure, $z_{\rm no}$ given in Eq.~(\ref{znocontrib}). To fix the value of $z$ in a more convenient way, we look at the most fundamental correlation function (\ref{Cdefperturbation}) obtained for $A=a_{\bm{k}}$ and $B=a^\dag_{\bm{k}'}$, that is, for a field quantum created with momentum $\bm{k}'$ and later annihilated with momentum $\bm{k}$. We introduce the correlation function $\Cco_{\bm{k}}(\omega^2)$ by
\begin{equation}\label{cpropadef}
    \frac{\Cco_{a_{\bm{k}} a^\dag_{\bm{k}'}}(-\omega)
    - \Cco_{a_{\bm{k}} a^\dag_{\bm{k}'}}(\omega)}{2\omega} =
    \Cco_{\bm{k}}(\omega^2) \, \delta(\bm{k}-\bm{k}') ,
\end{equation}
where momentum conservation has been taken into account and, in view of its symmetry in $\omega$, a dependence of $\Cco_{\bm{k}}$ on $\omega^2$ has been anticipated. By choosing $z$ such that the moment condition,
\begin{equation}\label{secmomcondz}
    \Cco_{\bm{0}}(\omega^2) \bigg|_{\omega^2=0} = m^2 \,
    \frac{\partial\Cco_{\bm{0}}(\omega^2)}{\partial\omega^2} \bigg|_{\omega^2=0}
\end{equation}
is fulfilled, we have a well-defined relationship between the mass parameter $m$ and a physical correlation function of the interacting system in the limit of small frequencies and zero wave vector. With this choice of $z$, the physical meaning of mass is not affected by interactions. This leads to a nicely organized perturbation theory in $\lambda$ that can be refined very conveniently by a renormalization-group analysis and avoids the usual need for mass renormalization. Allowing for $z \neq z_{\rm no}$ results in what is usually achieved by a counterterm. We assume that $z$ possesses a perturbation expansion in $\lambda$. It turns out below that this perturbation expansion of $z$ can be constructed very easily, in particular, if we rewrite it in the equivalent form
\begin{equation}\label{secmomcondzeq}
    \Cco_{\bm{0}}(0) = - \frac{1}{m^2} \, \frac{\partial}{\partial\omega^2}
    \, (\omega^2-m^2)^2 \, \Cco_{\bm{0}}(\omega^2) \bigg|_{\omega^2=0} .
\end{equation}

We further introduce the factor $Z$ by the condition
\begin{equation}\label{secmomcondZ}
    \Cco_{\bm{0}}(0) = Z \, \Cco_{\bm{0}}^{(0)}(0) ,
\end{equation}
so that we can preserve the normalization in the presence of interactions by rescaling the field operators or, equivalently, the creation and annihilation operators. Equation (\ref{secmomcondZ}) offers an explicit recipe for calculating $Z$. In particular, a perturbation theory for the two-time correlations on the left-hand side of Eq.~(\ref{secmomcondZ}) directly provides an explicit perturbation expansion for $Z$. Note that, in the thermodynamic approach, a change in the normalization of the fields in the Hamiltonian can trivially be absorbed in a change of the coupling constant $\lambda$ and an overall factor to be included into the definition of temperature.

\subsection{Friction mechanism}
We now apply the linearized thermodynamically consistent master equation (\ref{Grabertmelinrel}) to the treatment of quantum field theory. With the coupling operators
\begin{equation}\label{couplefieldtheory}
    Q_j, \, Q_k \mapsto \nabla^2 \varphi(\bm{x}), \, \nabla^2 \varphi(\bm{y}) ,
\end{equation}
which favor dissipation and smoothing of small-scale features through the occurrence of the Laplacian $\nabla^2$ as the simplest scalar differential operator, and the local relaxation rates
\begin{equation}\label{ratefieldtheory}
    \gamma_{jk} \mapsto 2 \gamma \, \delta(\bm{x}-\bm{y}) ,
\end{equation}
we arrive at the preliminary quantum master equation
\begin{eqnarray}
    \frac{d\hat{\rho}}{dt} &=& i \, \Qcommu{\hat{\rho}}{H} - \int d^dk \,
    \frac{\gamma_k}{\beta\omega_k} \nonumber \\
    &\times& \Kopi \Qcommu{{a_{\bm{k}}}^\dag + a_{-\bm{k}}}{\Kop \,
    \Qcommu{{a_{-\bm{k}}}^\dag + a_{\bm{k}}}{\hat{\rho}}} , \qquad
\label{QME}
\end{eqnarray}
where $\gamma_k = \gamma |\bm{k}|^4$ with the friction parameter $\gamma$. Note that the occurrence of $|\bm{k}|^4$ is natural because the scalar coupling operator, which involves the Laplacian, occurs quadratically in the quantum master equation.

The quantum master equation (\ref{QME}) is obtained under the assumption that only the Laplacian of the fields should be used for the dissipative coupling, not the Laplacian of their canonical conjugates. A more symmetric coupling to the fields and their conjugates leads to the simpler final equation
\begin{eqnarray}
    \frac{d\hat{\rho}}{dt} &=& \Lop\hat{\rho} = i \, \Qcommu{\hat{\rho}}{H}
    - \int d^dk \, \frac{\gamma_k}{\beta\omega_k} \nonumber \\
    &\times& \Kopi \Big( \Qcommu{{a_{\bm{k}}}^\dag}{\Kop \,
    \Qcommu{a_{\bm{k}}}{\hat{\rho}}} + \Qcommu{a_{\bm{k}}}{\Kop \,
    \Qcommu{{a_{\bm{k}}}^\dag}{\hat{\rho}}} \Big) . \qquad
\label{QMEsym}
\end{eqnarray}
Such a symmetric coupling also arises for the damping of electromagnetic field modes inside a cavity (see Eq.~(3.307) of \cite{BreuerPetru}). For single creation or annihilation operators, Eqs.~(\ref{condensmat}), (\ref{lnLemma}) and (\ref{acommutator}) lead to the remarkably simple results
\begin{equation}\label{Lprimesingle1}
    \Lop a^\dag_{\bm{k}} = - i \left( 1 - i \frac{\gamma_k}{\omega_k} \right)
    \Qcommu{H}{a^\dag_{\bm{k}}} ,
\end{equation}
and
\begin{equation}\label{Lprimesingle2}
    \bar{\Lop} a_{\bm{k}} = - i \left( 1 - i \frac{\gamma_k}{\omega_k} \right)
    \Qcommu{a_{\bm{k}}}{H} .
\end{equation}

The quantum master equation (\ref{QMEsym}) is our fundamental equation of quantum field theory obtained by a generalization of the canonical quantization procedure for dissipative systems. In addition to the usual reversible evolution given by the commutator with the Hamiltonian, local degrees of freedom are damped by an irreversible friction mechanism expressed in terms of a double commutator. The cutoff of a quantum field theory is realized as a spatial smoothing achieved by a dissipative dynamical mechanism where smoothing over short distances takes place very quickly.

\subsection{Free field theory}
In this section, we consider the calculation of equilibrium averages and the evolution equations for the free theory with $\lambda=0$. To generalize Wick's theorem for evaluating complicated moments to the case of finite temperatures, we use the ideas of the chapter ``field theory at finite temperature'' (Chap.~7) of \cite{FetterWalecka} (see also the theory of temperature Green's functions in Sec.~5.6 of \cite{KuboetalII}). Note that equilibrium averages are not affected by the fact that we use a quantum master equation with dissipation rather than a purely reversible Hamiltonian evolution.

In the spirit of Eqs.~(24.32) and (24.36) of \cite{FetterWalecka}, we obtain for averages performed with the free Hamiltonian $H^{(0)}$ for an arbitrary observable $A$:
\begin{equation}\label{Wick}
    \aves{a^\dag_{\bm{k}} A}^{(0)} =
    \frac{\aves{\Qcommu{A}{a^\dag_{\bm{k}}}}^{(0)}}{e^{\beta\omega_k}-1} \,, \quad
    \ave{a_{\bm{k}} A}^{(0)} =
    \frac{\ave{\Qcommu{a_{\bm{k}}}{A}}^{(0)}}{1-e^{-\beta\omega_k}} .
\end{equation}
Wick's theorem (\ref{Wick}) is the working horse for evaluating the free averages containing an increasing number of creation/annihilation operators and hence plays a crucial role in perturbation theory.

In the following, it is often convenient to consider observables of the normal ordered form,
\begin{equation}\label{normform}
    A = a^\dag_{\bm{k}'_1} \ldots a^\dag_{\bm{k}'_J}
    \, a_{\bm{k}_1} \ldots a_{\bm{k}_K} .
\end{equation}
In particular, we are interested in the free evolution of such a normal ordered operator $A$. We study the ingredients to the reversible and irreversible contributions in the evolution equation separately.

The \emph{reversible part} of the free evolution is obtained from the commutation relations
\begin{equation}\label{H0acommus}
    \Qcommu{H^{(0)}}{a_{\bm{k}}^\dag} = \omega_k a_{\bm{k}}^\dag \,, \qquad
    \Qcommu{H^{(0)}}{a_{\bm{k}}} = - \omega_k a_{\bm{k}} .
\end{equation}
These commutation relations imply the more general result
\begin{equation}\label{H0Anocommu}
    \Qcommu{H^{(0)}}{A} = \omega_A A ,
\end{equation}
where the frequency
\begin{equation}\label{omegaAdef}
    \omega_A = \sum_{j=1}^J \omega_{k'_j} - \sum_{j=1}^K \omega_{k_j}
\end{equation}
has been associated with the operator $A$.

By following the change with $u$, we obtain as a useful consequence of Eq.~(\ref{H0Anocommu})
\begin{equation}\label{expH0Anocommu}
    e^{-u H^{(0)}} A \, e^{u H^{(0)}} = e^{-u \omega_A} \, A ,
\end{equation}
for arbitrary complex $u$. The normal ordering of the operators in Eq.~(\ref{normform}) is actually irrelevant for the above results (\ref{H0Anocommu}) and (\ref{expH0Anocommu}). We further obtain
\begin{equation}\label{K0A}
    \Kop^{(0)} A = w(\omega_A) \, A \rho^{(0)} = w(-\omega_A) \, \rho^{(0)} A ,
\end{equation}
where $\rho^{(0)}$ is the equilibrium density matrix (\ref{eqsolqme}) for the free field theory and the function $w(\omega)$ is defined as
\begin{equation}\label{womdef}
    w(\omega) = \frac{1 - e^{- \beta\omega}}{\beta\omega} .
\end{equation}
Equation (\ref{K0A}) follows from Eqs.~(\ref{condensmat}) and (\ref{expH0Anocommu}) by explicit integration over $u$. From Eq.~(\ref{K0A}) we further obtain
\begin{equation}\label{cancorfreetheo}
    \cancor{A}{B}^{(0)} = w(-\omega_A) \, \ave{A B}^{(0)} = w(\omega_B) \, \ave{A B}^{(0)} ,
\end{equation}
and, with the additional help of Eq.~(\ref{expH0Anocommu}),
\begin{equation}\label{comavefreetheo}
    \ave{\Qcommu{A}{B}}^{(0)} = - \beta\omega_A \cancor{A}{B}^{(0)}
    = \beta\omega_B \cancor{A}{B}^{(0)} ,
\end{equation}
so that all kinds of free equilibrium averages are related by simple factors. Note that Eq.~(\ref{comavefreetheo}) can also be obtained as a simple special case of the much more general result (\ref{comcangenid}) for the interacting system.

The \emph{irreversible part} of the evolution super-operator of the free theory can now be evaluated by means of the identity (\ref{K0A}). As a first step, we obtain the useful identities
\begin{eqnarray}
    && \hspace{-2em} {\Kop^{(0)}}^{-1} \Qcommu{A}{\Kop^{(0)}B} = \nonumber \\
    && \beta \omega_A \frac{w(\omega_A) \, w(\omega_B)}{w(\omega_A+\omega_B)} BA
    + \frac{w(\omega_B)}{w(\omega_A+\omega_B)} \Qcommu{A}{B} = \nonumber \\
    && \beta \omega_A \frac{w(\omega_A) \, w(\omega_B)}{w(\omega_A+\omega_B)} AB
    + \frac{w(\omega_B) \, e^{- \beta\omega_A}}{w(\omega_A+\omega_B)} \Qcommu{A}{B} . \qquad
\label{K0Acor}
\end{eqnarray}
After using Eqs.~(\ref{H0Anocommu}) and (\ref{K0Acor}) in Eq.~(\ref{QMEsym}), the total free evolution super-operator resulting as the sum of the reversible and irreversible contributions is found to be
\begin{eqnarray}
    \Lop^{(0)} A &=& - i \omega_A A - \int d^dk \, \frac{\gamma_k}{w(\omega_A)} \nonumber \\
    &\times& \bigg\{
    w(\omega_A - \omega_k) \, w(\omega_k) \, {a_{\bm{k}}}^\dag \Qcommu{a_{\bm{k}}}{A}
    \nonumber \\
    && + \, w(\omega_A + \omega_k) \, w(-\omega_k) \, \Qcommu{A}{{a_{\bm{k}}}^\dag} a_{\bm{k}}
    \nonumber \\
    && - \, \frac{e^{-\beta\omega_k}}{\beta\omega_k}
    w(\omega_A - \omega_k) \, \Qcommu{\Qcommu{a_{\bm{k}}}{A}}{{a_{\bm{k}}}^\dag}
    \nonumber \\
    && - \, \frac{1}{\beta\omega_k} w(\omega_A + \omega_k) \,
    \Qcommu{a_{\bm{k}}}{\Qcommu{A}{{a_{\bm{k}}}^\dag}} \bigg\} . \qquad
\label{freeevol}
\end{eqnarray}
Note that the double commutators in the last two terms in Eq.~(\ref{freeevol}) are actually equal. According to the first two terms under the integral in Eq.~(\ref{freeevol}), the decay rate $\gamma_A$ for an operator $A$ is given by
\begin{eqnarray}
    \gamma_A &=& \sum_{j=1}^J \gamma_{k'_j} W(\omega_{k'_j},\omega_A-\omega_{k'_j})
    \nonumber \\
    &+& \sum_{j=1}^K \gamma_{k_j} W(-\omega_{k_j},\omega_A+\omega_{k_j}) ,
\label{gammaAdef}
\end{eqnarray}
with
\begin{equation}\label{WAomdef}
    W(\omega,\omega') = \frac{w(\omega) \, w(\omega')}{w(\omega+\omega')} .
\end{equation}
The function $W$ has the useful symmetry properties,
\begin{equation}\label{Wsymproper}
    W(\omega,\omega') = W(\omega',\omega) , \quad
    W(\omega,\omega') = W(-\omega,-\omega') .
\end{equation}
By simplifying the result (\ref{freeevol}) for normal-ordered products (\ref{normform}) of creation and annihilation operators, the fundamental evolution operator of the free theory can now be rewritten in the compact form
\begin{equation}\label{freeevols}
    \Lop^{(0)} A = - (i \omega_A + \gamma_A) A + \Gamma_A ,
\end{equation}
where the term $\Gamma_A$ consists of all the contributions obtained by deleting one creation and one annihilation operator from $A$. The explicit normal ordered form of $\Gamma_A$ is given by
\begin{eqnarray}
  \Gamma_A &=& \int d^dk \, \frac{\gamma_k}{\beta\omega_k}
  \frac{1}{w(\omega_A)} \Big\{ w(\omega_A + \omega_k) \nonumber \\
  && \qquad + \, e^{-\beta\omega_k} \, w(\omega_A - \omega_k) \Big\} \,
  \Qcommu{a_{\bm{k}}}{\Qcommu{A}{a^\dag_{\bm{k}}}} . \qquad
\label{GammaAexpl}
\end{eqnarray}
When exponentially small terms are neglected in the low-temperature limit, Eq.~(\ref{GammaAexpl}) for $\omega_A \neq 0$ can be simplified to
\begin{equation}\label{GammaAexpls}
  \Gamma_A = \int d^dk \, \frac{\gamma_k \, |\omega_A|}{\beta\omega_k(\omega_k+|\omega_A|)}
  \, \Qcommu{a_{\bm{k}}}{\Qcommu{A}{a^\dag_{\bm{k}}}} .
\end{equation}
Some examples of $\Gamma_A$ for normal ordered products of up to three creation/annihilation operators are given in Appendix~\ref{app_GammaAexamples}.

From the identity (\ref{L0R0ident}) and the explicit expression (\ref{freeevols}), we obtain the result
\begin{equation}\label{R0reduction}
    \Ropz(\omega) A = \frac{1}{i \omega + i \omega_A + \gamma_A} \,
    \left[ A + \Ropz(\omega) \Gamma_A \right] ,
\end{equation}
by which the calculation of $\Ropz(\omega) A$ can be reduced to successively simpler products of creation and annihilation operators. We similarly have
\begin{equation}\label{R0barreduction}
    \bar{\Rop}^{(0)}(\omega) A = \frac{1}{i \omega - i \omega_A + \gamma_A} \,
    \left[ A + \bar{\Rop}^{(0)}(\omega) \Gamma_A \right] .
\end{equation}

\subsection{Propagator}\label{subsecprop}
We are now ready to analyze the fundamental correlation function $\Cco_{\bm{k}}(\omega^2)$ introduced in Eq.~(\ref{cpropadef}) for an interacting system. If we introduce the equilibrium averages
\begin{equation}\label{pertXkdef}
    \ave{\Qcommu{\Qcommu{a_{\bm{k}}}{H^{(1)}}}{a^\dag_{\bm{k}'}}} =
    X_{\bm{k}} \, \delta(\bm{k}-\bm{k}') ,
\end{equation}
and
\begin{equation}\label{pertYkdef}
    \ave{\Qcommu{\bar{\Rop}^{(0)}(\omega)\Qcommu{a_{\bm{k}}}{H^{(1)}}}{\Qcommu{H^{(1)}}{a^\dag_{\bm{k}'}}}} =
    i \, Y_{\bm{k}}(\omega) \, \delta(\bm{k}-\bm{k}') ,
\end{equation}
and use the results (\ref{Lprimesingle1}), (\ref{Lprimesingle2}), (\ref{R0reduction}), and (\ref{R0barreduction}), then Eqs.~(\ref{Cdefperturbation}) and (\ref{cpropadef}) lead to
\begin{eqnarray}
    && \hspace{-1em} \Cco_{\bm{k}}(\omega^2) = \frac{i}{\omega^2 - (\omega_k-i\gamma_k)^2}
    \nonumber \\
    && + \, \frac{2 i (\omega_k - i\gamma_k)^2}{\omega_k^2 [\omega^2 - (\omega_k-i\gamma_k)^2]^2}
    \, \Big[ \omega_k X_{\bm{k}} + (\omega_k-i\gamma_k) Y^+_{\bm{k}}(\omega) \Big]
    \nonumber \\
    && - \, \frac{i (\omega_k - i\gamma_k)^2 [\omega^2 + (\omega_k-i\gamma_k)^2]}{
    \omega_k^2 [\omega^2 - (\omega_k-i\gamma_k)^2]^2}
    \, \frac{Y^-_{\bm{k}}(\omega)}{\omega} ,
\label{fundpropres}
\end{eqnarray}
where $Y^+_{\bm{k}}(\omega)$ and $Y^-_{\bm{k}}(\omega)$ are the symmetric and antisymmetric contributions to $Y_{\bm{k}}(\omega)$.

The exact results for $X_{\bm{k}}$ and $Y_{\bm{k}}(\omega)$ for any temperature are given in Appendix~\ref{app_exact2ndord}. Equations (\ref{X2full}) and (\ref{Y2full}) allow us to see explicitly how the limits of low temperature and weak friction can be performed. Neglecting exponentially small terms at low temperatures, we find
\begin{equation}\label{X2limit}
    X_{\bm{k}} = \frac{\lambda z}{\tilde{\omega}_k} ,
\end{equation}
and
\begin{eqnarray}
    Y_{\bm{k}}(\omega) &=& \frac{2\lambda^2 z^2}{\tilde{\omega}_k^2}
    \frac{\omega_k}{(\omega-i\gamma_k)^2 - \omega_k^2}
    \nonumber \\
    && \hspace{-3em} - \, \frac{\lambda^2}{96\,\tilde{\omega}_k} \frac{1}{(2\pi)^{2d}}
    \int \frac{d^dk_1 d^dk_2 d^dk_3}{\tilde{\omega}_{k_1}
    \tilde{\omega}_{k_2} \tilde{\omega}_{k_3}} \,
    \delta(\bm{k}_1+\bm{k}_2+\bm{k}_3+\bm{k})
    \nonumber \\
    &\times& r(\omega, \omega_{k_1}+\omega_{k_2}+\omega_{k_3}, \gamma_{k_1 k_2 k_3}) ,
\label{Y2limit}
\end{eqnarray}
with the damping rate
\begin{eqnarray}
    \gamma_{k_1 k_2 k_3} &=& \frac{\omega_{k_1}+\omega_{k_2}+\omega_{k_3}}{\beta}
    \bigg[ \frac{\gamma_{k_1}}{\omega_{k_1}(\omega_{k_2}+\omega_{k_3})} \nonumber \\
    &+& \frac{\gamma_{k_2}}{\omega_{k_2}(\omega_{k_1}+\omega_{k_3})}
    + \frac{\gamma_{k_3}}{\omega_{k_3}(\omega_{k_1}+\omega_{k_2})} \bigg] , \qquad
\label{gammatripdef}
\end{eqnarray}
and the rational kernel function
\begin{equation}\label{rkernfunc}
    r(\omega, \bar{\omega}, \bar{\gamma}) =
    \frac{1}{\omega+\bar{\omega}- i \bar{\gamma}}
    - \frac{1}{\omega-\bar{\omega}- i \bar{\gamma}} .
\end{equation}

The occurrence of $i \bar{\gamma}$ in Eq.~(\ref{rkernfunc}) implies specific rules for the treatment of the poles in the integrations of Eq.~(\ref{Y2limit}). The $i \bar{\gamma}$ hence plays a similar role as the $i \varepsilon$ in the usual approach (see, for example, Eqs.~(\ref{IntegralI1}), (\ref{Integralgenexp2}) and the comments offered there). The usual $i \varepsilon$ resolves infrared problems by clarifying causality issues. In our approach based on irreversible equations, there is a natural arrow of time and a large scale decay of correlations that leads to convergent integrals. Therefore, the small friction coefficient simultaneously resolves infrared problems by ensuring a decay of correlations on large scales and ultraviolet problems by dissipative smoothing of the fields on small scales.

\begin{figure}
\centerline{\epsfysize=3.5cm \epsffile{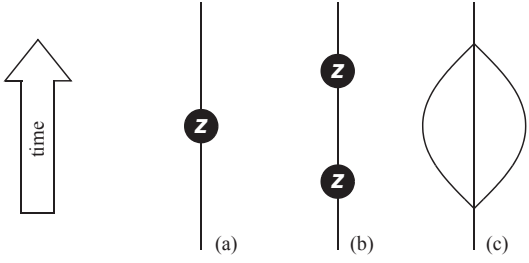}} \caption[ ]{Feynman diagrams potentially contributing to the propagator in second-order perturbation theory.} \label{fig_feynman_diagr2}
\end{figure}

As the averages in Eqs.~(\ref{pertXkdef}) and (\ref{pertYkdef}) need to be evaluated by means of Wick's theorem, we can introduce Feynman diagrams to keep track of the contraction structure of the various contributions. The contribution $X_k$ corresponds to the diagram in Fig.~\ref{fig_feynman_diagr2}(a), whereas the contribution $Y_{\bm{k}}(\omega)$ consists of two terms represented by the Feynman diagrams in Figs.~\ref{fig_feynman_diagr2}(b) and (c). The diagram in Fig.~\ref{fig_feynman_diagr2}(a) looks like a first-order contribution, as also the factor of $\lambda$ in Eq.~(\ref{X2limit}) suggests. Note, however, that this contribution stems from the quadratic part of the interaction Hamiltonian (\ref{H1defNO}) arising in the normal-ordering procedure and that it hence contains a factor of $z$, which is shown as an important reminder in the Feynman diagrams. This factor $z$ still needs to be determined from Eq.~(\ref{secmomcondz}) or (\ref{secmomcondzeq}). It turns out that $z$ itself is of order $\lambda$ so that the diagrams in Figs.~\ref{fig_feynman_diagr2}(a) and (b) actually represent second and fourth order contributions, respectively. The contribution in Fig.~\ref{fig_feynman_diagr2}(c) clearly is a genuine second-order term as suggested by the factor $\lambda^2$ in front of the triple momentum integral in Eq.~(\ref{Y2limit}).

In the limit of vanishing friction, we can set $\gamma_k = 0$ in Eqs.~(\ref{fundpropres}) and (\ref{Y2limit}). Equation (\ref{fundpropres}) becomes
\begin{eqnarray}
    (\omega^2 - \omega_k^2)^2 \, \Cco_{\bm{k}}(\omega^2) &=& i(\omega^2 - \omega_k^2)
    + 2 i \omega_k \Big[ X_{\bm{k}} + Y^+_{\bm{k}}(\omega) \Big]
    \nonumber \\
    &-& \, i (\omega^2 + \omega_k^2) \, \frac{Y^-_{\bm{k}}(\omega)}{\omega} .
\label{fundpropresx}
\end{eqnarray}
In simplifying Eq.~(\ref{Y2limit}) we need to be more careful with $\gamma_{k_1 k_2 k_3}$ because this quantity keeps all integrals finite, which expresses the regularizing character of the smoothing friction mechanism. To simplify the kernel function occurring in Eq.~(\ref{Y2limit}), we consider the following expansion in $\omega$,
\begin{equation}\label{rkernexpan1}
    r(\omega, \bar{\omega}, \bar{\gamma}) = \sum_{n=0}^\infty
    \left[ \frac{1}{(\bar{\omega}+i\bar{\gamma})^{n+1}}
    + \frac{(-1)^n}{(\bar{\omega}-i\bar{\gamma})^{n+1}} \right] \omega^n .
\end{equation}
After neglecting the regularizing friction mechanism in all convergent integrals (where, for simplicity, we assume $d < 3$, so that the integrals converge for $n \geq 2$), we obtain the simplified expansion
\begin{eqnarray}
    r(\omega, \bar{\omega}, \bar{\gamma}) &=&
    \frac{1}{\bar{\omega}+i\bar{\gamma}}
    + \frac{1}{\bar{\omega}-i\bar{\gamma}}
    + \frac{\omega}{(\bar{\omega}+i\bar{\gamma})^2}
    - \frac{\omega}{(\bar{\omega}-i\bar{\gamma})^2}
    \nonumber \\
    &+& 2 \sum_{n=1}^\infty \frac{\omega^{2n}}{\bar{\omega}^{2n+1}} .
\label{rkernexpan2}
\end{eqnarray}
We next analyze the terms proportional to $\omega$ in Eq.~(\ref{rkernexpan2}). Note from Eq.~(\ref{gammatripdef}) that the regularizing parameter is $\gamma/\beta$, and dimensional analysis hence implies that the corresponding smoothing length scale is given by
\begin{equation}\label{smoothlength}
    \ell = \sqrt{\frac{2 \gamma}{\beta}} ,
\end{equation}
where the factor $2$ is introduced for later convenience. Each of the resulting two contributions to $Y_{\bm{k}}$ proportional to $\omega$ diverges as $\ell^{-(2d-5)}$ (for $d<5/2$, the terms are convergent and cancel each other in the limit of zero friction). For symmetry reasons, the divergent terms (for $5/2<d<3$) cancel each other and the corrections, when expanded in terms of the dimensionless parameter $\ell m$, vanish in the limit of small friction. We can hence rewrite the kernel function (\ref{rkernexpan2}) as
\begin{equation}\label{rkernexpan3}
    r(\omega, \bar{\omega}, \bar{\gamma}) =
    \frac{2 \bar{\omega}}{\bar{\omega}^2+\bar{\gamma}^2}
    + 2 \sum_{n=1}^\infty \frac{\omega^{2n}}{\bar{\omega}^{2n+1}} ,
\end{equation}
implying $Y^+_{\bm{k}}(\omega) = Y_{\bm{k}}(\omega)$ and $Y^-_{\bm{k}}(\omega) = 0$ for the symmetric and antisymmetric contributions to $Y_{\bm{k}}(\omega)$. At the expense of losing the convergence of the integral of each individual term, Eq.~(\ref{rkernexpan3}) can formally be re-summed into a simple rational function of $\omega^2$ and a constant term with prefactor $\bar{\gamma}^2$,
\begin{equation}\label{rkernexpan4}
    r(\omega, \bar{\omega}, \bar{\gamma}) =
    \frac{2 \bar{\omega}}{\bar{\omega}^2-\omega^2}
    - \frac{2 \bar{\gamma}^2}{\bar{\omega}(\bar{\omega}^2+\bar{\gamma}^2)} .
\end{equation}

After simplifying the kernel function, we are now in a position to discuss the propagator in more detail. For the second-order expansion $Y_{\bm{0}}(\omega) = Y + Y' \omega^2$, the conditions (\ref{secmomcondzeq}) and (\ref{secmomcondZ}) for the correlation function (\ref{fundpropresx}) imply
\begin{equation}\label{secmomcondcr1}
    X_{\bm{0}} = - ( Y + m^2 Y') ,
\end{equation}
and
\begin{equation}\label{secmomcondcr2}
    Z = 1 + 2 m Y' .
\end{equation}
Note that the value of $z$ following from Eqs.~(\ref{secmomcondcr1}), (\ref{X2limit}), and (\ref{Y2limit}) with the kernel function (\ref{rkernexpan3}), for which we obtain the straightforward result
\begin{eqnarray}
    z &=& \frac{\lambda}{48} \, \frac{1}{(2\pi)^{2d}}
    \int \frac{d^dk_1 d^dk_2 d^dk_3}{\tilde{\omega}_{k_1}
    \tilde{\omega}_{k_2} \tilde{\omega}_{k_3}} \,
    \frac{\delta(\bm{k}_1+\bm{k}_2+\bm{k}_3)}{\omega_{k_1}+\omega_{k_2}+\omega_{k_3}}
    \nonumber\\
    &\times& \frac{(\omega_{k_1}+\omega_{k_2}+\omega_{k_3})^2 + m^2}{
    (\omega_{k_1}+\omega_{k_2}+\omega_{k_3})^2 + \gamma_{k_1 k_2 k_3}^2} ,
\label{secmomsolz}
\end{eqnarray}
turns out to be small, more precisely, only of first order in $\lambda$. In terms of the integrals introduced in \cite{ZinnJustin} and reproduced in Eq.~(\ref{IntegralI2I3}), we find
\begin{equation}\label{secmomcondcr3}
    Z = 1 + \frac{1}{6} \, (\lambda m^{d-3})^2 \, I_3 ,
\end{equation}
and
\begin{equation}\label{secmomsolzz}
    z = \frac{\lambda}{12} \, m^{2d-4} \, (I_2-I_3) .
\end{equation}
The result (\ref{secmomcondcr3}) for the normalization factor $Z$ coincides with the one given in Eqs.~(11.64) and (11.69) of \cite{ZinnJustin}. The quantity $z$ is usually taken into account through mass renormalization. More precisely, we have verified explicitly that $Z \, [1 + 2 \lambda \, (z-z_{\rm no})/m^2]$ coincides with the mass renormalization factor $Z_m$ of \cite{ZinnJustin}.

For presenting the results of the perturbation theory, it is convenient to ``amputate the external legs'' associated with free propagators (see, for example, p.~55 of \cite{Zee}). We hence introduce the usual amplitude
\begin{equation}\label{amplitudedef}
    \Mco_{\bm{k}}(\omega^2) = - \frac{1}{Z} \,
    (\omega^2 - \omega_k^2)^2 \, \Cco_{\bm{k}}(\omega^2) .
\end{equation}
From Eqs.~(\ref{fundpropresx}) and (\ref{secmomcondcr2}), we obtain for the second-order perturbation series
\begin{equation}\label{amplitudeper}
    \Mco_{\bm{k}}(\omega^2) = - i(\omega^2 - \omega_k^2)
    - 2 i \Big[ \omega_k Y_{\bm{k}}(\omega) - m Y - (\omega^2 - \bm{k}^2) Y' \Big]
\end{equation}
with nicely Lorentz invariant correction terms due to the constants $Y$ and $Y'$. The Lorentz invariance of
\begin{eqnarray}
    \omega_k Y_{\bm{k}}(\omega) &=&  - \, \frac{\lambda^2}{96} \frac{1}{(2\pi)^{2d}}
    \int \frac{d^dk_1 d^dk_2 d^dk_3}{\tilde{\omega}_{k_1}
    \tilde{\omega}_{k_2} \tilde{\omega}_{k_3}}
    \nonumber \\
    &\times& \delta(\bm{k}_1+\bm{k}_2+\bm{k}_3+\bm{k})
    \nonumber \\
    &\times& r(\omega, \omega_{k_1}+\omega_{k_2}+\omega_{k_3}, \gamma_{k_1 k_2 k_3}) , \qquad
\label{Y2limitx}
\end{eqnarray}
is less obvious. Note that we have assumed $\tilde{\omega}_k = \omega_k$ for the externally fixed momentum (regularization is only relevant in integrals). To establish the Lorentz invariance of $\omega_k Y_{\bm{k}}(\omega)$, we need to pass from space to spacetime integrals. If we neglect all regularization effects by assuming also $\tilde{\omega}_{k_j} = \omega_{k_j}$ and by considering the limit of vanishing friction,
\begin{equation}\label{rkernexpan4lim}
    r(\omega, \bar{\omega}, 0) =
    \frac{2 \bar{\omega}}{\bar{\omega}^2-\omega^2} ,
\end{equation}
we find $\omega_k Y_{\bm{k}}(\omega) = \lambda^2 I_{\bm{k}}(\omega^2)/12$ in terms of the integral defined in Eq.~(\ref{Integralgenexp1}). Equivalent forms of the integral $I_{\bm{k}}(\omega^2)$ are given in Appendix~\ref{app_integrals}. The representation in Eq.~(\ref{Integralgenexp2}) demonstrates the invariance of $\omega_k Y_{\bm{k}}(\omega)$, clearly exhibits the structure of a product of three free spacetime propagators, and reproduces the standard result for $\Mco_{\bm{k}}(\omega^2)$ as, for example, given in Eq.~(III.3.2) of \cite{Zee}. Equation (\ref{Integralgenexp3}) finally leads us to the Eucledian time version of the second-order perturbation series for the propagator given in Eqs.~(11.36) and (11.38) of \cite{ZinnJustin}.

Regularization mechanisms often destroy the Lorentz invariance, although a covariant formulation of friction within a not manifestly covariant thermodynamic framework may be possible (see, for example, \cite{hco109,hco111,hco112}). After neglecting the regularization mechanism for establishing Lorentz invariance, we next look at Eq.~(\ref{secmomsolz}) or at Eq.~(\ref{Y2limitx}) with the kernel function (\ref{rkernexpan3}) to see how integrals become regularized. In the traditional approach, one has $\gamma_{k_1 k_2 k_3} = 0$ and the convergence of integrals is achieved by an enhanced increase of the frequencies $\tilde{\omega}_{k_j}$ at large wave vectors $\bm{k}_j$. In the present approach, we can use $\tilde{\omega}_{k_j} = \omega_{k_j}$ because the occurrence of $\gamma_{k_1 k_2 k_3}^2$ keeps all integrals finite.

\subsection{Four-point correlation}\label{subsecfourpt}
Our next goal is to obtain the $\beta$ function for the coupling constant from a second-order perturbation theory of the form (\ref{perturbationform}). As a first-order term is missing in the perturbation series (\ref{amplitudeper}) for the propagator, we cannot determine the fixed-point value $\lambda^*$ of the dimensionless coupling constant. We hence consider a four-point correlation which, to leading order, is given by the interaction strength $\lambda$ of $\varphi^4$ theory.

As a second example, we hence apply our basic perturbation formula (\ref{Cdefperturbation}) to $A=a_{\bm{k}_1} a_{\bm{k}_2}$ and $B=a^\dag_{\bm{k}'_1} a^\dag_{\bm{k}'_2}$. The result is of little direct interest because two particles are created at exactly the same time and later annihilated also at exactly the same time. Nevertheless, we will be able to produce a perturbation series from which we can read off the parameter $\lambda^*$ and the $\beta$ function (\ref{beta2nd}). It is actually sufficient to do the calculations in the limit of vanishing frequency $\omega$ and wave vectors $\bm{k}_j$, $\bm{k}'_j$.

According to Eq.~(\ref{R0barreduction}), the zeroth-order contribution in Eq.~(\ref{Cdefperturbation}) factorizes into two propagators. After subtracting a suitable product of propagators, we are interested only in contributions corresponding to connected Feynman diagrams. As we restrict ourselves to the limits of vanishing frequency and friction (except for the regularization effect in integrals), we consider the property
\begin{equation}\label{Gammadef}
    \Gamma = \ave{\Qcommu{\bar{\Lop}'_{\rm rev} A}{B}}
    + \ave{\Qcommu{\bar{\Rop}^{(0)}(0) \bar{\Lop}'_{\rm rev} A}{\Lop'_{\rm rev} B}} .
\end{equation}
With the explicit form of $\Qcommu{\Qcommu{A}{H^{(1)}}}{B}$ given in Eq.~(\ref{H1_dcommu_aaadadx}), we obtain
\begin{equation}\label{perturbationfoc2a}
    \ave{\Qcommu{\bar{\Lop}'_{\rm rev} A}{B}} =
    - i \ave{\Qcommu{\Qcommu{A}{H^{(1)}}}{B}} = - i \lambda \, \frac{\cal F}{4} ,
\end{equation}
with
\begin{equation}\label{perturbationfoc2b}
    {\cal F} = \frac{1}{(2\pi)^d}
    \frac{\delta(\bm{k}_1+\bm{k}_2-\bm{k}'_1-\bm{k}'_2)}{\sqrt{\tilde{\omega}_{k_1}
    \tilde{\omega}_{k_2} \tilde{\omega}_{k'_1} \tilde{\omega}_{k'_2}}} ,
\end{equation}
where we have omitted all terms that vanish exponentially at low temperatures and all products of two propagators corresponding to particles that do not interact with each other. Note that no second-order corrections resulting from the expansion (\ref{averageperturb}) have survived in Eq.~(\ref{perturbationfoc2a}).

\begin{figure}
\centerline{\epsfysize=3.5cm \epsffile{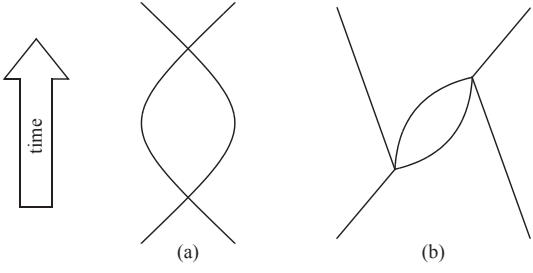}} \caption[ ]{Feynman diagrams contributing to the interaction vortex in second-order perturbation theory.} \label{fig_feynman_diagr4}
\end{figure}

According to Eqs.~(\ref{H1_commu_aa}) and (\ref{H1_commu_adad}), there are two types of contributions associated with the second-order term in Eq.~(\ref{Gammadef}). They can be represented by the Feynman diagrams shown in Fig.~\ref{fig_feynman_diagr4}. For the contribution associated with Fig.~\ref{fig_feynman_diagr4}(a), we have the explicit formula
\begin{eqnarray}
    && \hspace{-2em} \ave{\QcommuB{\bar{\Rop}^{(0)}(0) \,
    \Qcommu{a_{\bm{k}_1}}{\Qcommu{a_{\bm{k}_2}}{H^{(1)}}}}{
    \Qcommu{\Qcommu{H^{(1)}}{a^\dag_{\bm{k}'_1}}}{a^\dag_{\bm{k}'_2}}}} = \nonumber\\
    && \frac{\lambda^2}{32} \, \frac{{\cal F}}{(2\pi)^d}
    \int \frac{d^dq_1 d^dq_2}{\tilde{\omega}_{q_1} \tilde{\omega}_{q_2}} \,
    \delta(\bm{k}_1+\bm{k}_2+\bm{q}_1+\bm{q}_2) \nonumber\\
    && \times \, \Bigg [ \frac{1}{i(\omega_{q_1}
    +\omega_{q_2}) + \gamma_{a_{-\bm{q}_1} a_{-\bm{q}_2}}} \nonumber\\
    && \hspace{5em} + \, \frac{1}{i(\omega_{q_1}
    +\omega_{q_2}) - \gamma_{a^\dag_{\bm{q}_1} a^\dag_{\bm{q}_2}}} \Bigg ] ,
\label{Gammaperturbation1}
\end{eqnarray}
where the expressions (\ref{H1_dcommu_aa}), (\ref{H1_dcommu_adad}) and Wick's theorem (\ref{Wick}) have been used to arrive at this integral expression. A typical contribution associated with Fig.~\ref{fig_feynman_diagr4}(b) is given by
\begin{eqnarray}
    \ave{\QcommuB{\bar{\Rop}^{(0)}(0) \, \Qcommu{a_{\bm{k}_1}}{H^{(1)}}
    \, a_{\bm{k}_2}}{a^\dag_{\bm{k}'_1} \, \Qcommu{H^{(1)}}{a^\dag_{\bm{k}'_2}}}} =
    \hspace{3em} && \nonumber\\
    \frac{\lambda^2}{32} \, \frac{{\cal F}}{(2\pi)^d}
    \int \frac{d^dq_1 d^dq_2}{\tilde{\omega}_{q_1} \tilde{\omega}_{q_2}}
    \hspace{11.5em} && \nonumber\\ \times \,
    \frac{\delta(\bm{q}_1+\bm{q}_2+\bm{k}_1-\bm{k}'_1)}{i(\omega_{q_1}
    +\omega_{q_2}+\omega_{k'_1}+\omega_{k_2}) + \gamma_{a_{-\bm{q}_1}
    a_{-\bm{q}_2} a_{\bm{k}'_1} a_{\bm{k}_2}}} , \quad &&
\label{Gammaperturbation2}
\end{eqnarray}
where the evaluation is based on Eqs.~(\ref{H1_commu_a}) and (\ref{H1_commu_ad}). In total, there are four contributions of the type (\ref{Gammaperturbation2}) because $\bm{k}_1$ and $\bm{k}_2$ can be exchanged, as well as $\bm{k}'_1$ and $\bm{k}'_2$. All other terms are exponentially small.

Upon setting $\bm{k}_1 = \bm{k}_2 = \bm{k}'_1 = \bm{k}'_2 = \bm{0}$ in the integrals of Eqs.~(\ref{Gammaperturbation1}) and (\ref{Gammaperturbation2}) and inserting them into Eq.~(\ref{Gammadef}), we obtain the simple perturbation expansion
\begin{eqnarray}
    {\rm Im} \, \Gamma &=& \frac{\cal F}{4} \Bigg [ - \lambda
    + \frac{\lambda^2}{4} \, \frac{1}{(2\pi)^d}
    \int \frac{d^dq}{2 \tilde{\omega}_q^2 \omega_q (1+{\gamma'_q}^2)} \qquad \nonumber \\
    &+& \frac{\lambda^2}{2} \, \frac{1}{(2\pi)^d}
    \int \frac{d^dq}{2 \tilde{\omega}_q^2 (\omega_q+m) (1+{\gamma''_q}^2)} \Bigg ] ,
\label{vortexfinalcom}
\end{eqnarray}
with
\begin{equation}\label{gammadpdef}
    \gamma'_q = \frac{2 \gamma q^4}{\beta \omega_q^2} , \qquad
    \gamma''_q = \frac{2 \gamma q^4}{\beta \omega_q (\omega_q+2m)} .
\end{equation}
In evaluating $\gamma_{a_{-\bm{q}_1} a_{-\bm{q}_2}}$, $\gamma_{a^\dag_{\bm{q}_1} a^\dag_{\bm{q}_2}}$, and $\gamma_{a_{-\bm{q}_1} a_{-\bm{q}_2} a_{\bm{k}'_1} a_{\bm{k}_2}}$, we have once more neglected exponentially small terms at low temperatures. In the definition of the four-point correlation, we have not included a factor $1/Z^2$ because it would affect only third-order terms.

\begin{figure}
\centerline{\epsfxsize=6.5cm \epsffile{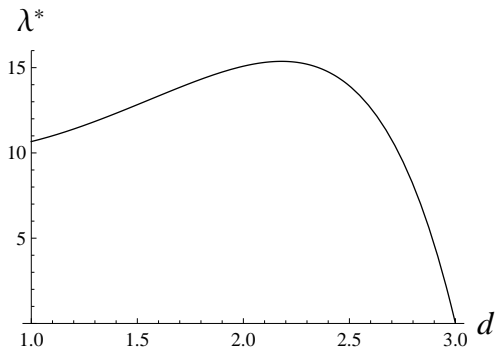}} \caption[ ]{Fixed point value $\lambda^*$ of the dimensionless coupling constant resulting from second-order perturbation theory as a function of space dimensionality $d$.} \label{fig_fixed_point}
\end{figure}

For $d<3$, the integrals in Eq.~(\ref{vortexfinalcom}) are nicely convergent without any regularization. For $\tilde{\omega}_q = \omega_q$ and $\gamma = 0$, these integrals have been expressed in terms of the $\Gamma$ function in Eqs.~(\ref{IntegralI1a}) and (\ref{IntegralI1b}). By evaluating the leading-order corrections resulting from small values of the friction parameter $\gamma$, we arrive at
\begin{eqnarray}
    {\rm Im} \, \Gamma &=& \frac{\cal F}{4} \Bigg [ - \lambda
    + \frac{1}{4} \lambda^2 \, m^{d-3} \, ( I'_1 + 2 I''_1 ) \nonumber \\
    &-&  \frac{3}{8} \lambda^2 \, \ell^{3-d} \, \frac{1}{(2\pi)^d}
    \int \frac{q \, d^dq}{1 + q^4} \Bigg ] ,
\label{vortexfinalcomx}
\end{eqnarray}
where the smoothing length scale $\ell$ has been defined in Eq.~(\ref{smoothlength}). We finally have a perturbation expansion with a first-order term to compare to the general form (\ref{perturbationform}) and we find a perfect match in structure. After using Eq.~(\ref{dintto1int}), the remaining one-dimensional integral can be evaluated in closed form (for example, with 3.241.2 of \cite{GradshteynRyzhik}). From the comparison of the general form (\ref{perturbationform}) with our perturbation expansion (\ref{vortexfinalcomx}), we then obtain the exponents
\begin{equation}\label{alpharesult}
    \alpha = \epsilon = 3 - d ,
\end{equation}
and the fixed point value
\begin{equation}\label{wow}
    \lambda^* = \frac{16}{3} \, 2^d \, \Gamma\left(\frac{d}{2}\right) \pi^{\frac{d-2}{2}}
    \, \sin \left[ (3-d) \frac{\pi}{4} \right]
    \approx \frac{16}{3} \pi^2 \epsilon .
\end{equation}
In one and two space dimensions, we obtain $\lambda^* = 32 / 3$ and $\lambda^* = 32 \sqrt{2} / 3$, respectively; in three space dimensions, we recover the free theory on large scales. The nontrivial dependence of $\lambda^*$ on the space dimensionality $d$, which is deeply related to the dissipative smoothing mechanism, is shown in Fig.~\ref{fig_fixed_point}. Note that the occurrence of $q^4$ in the denominator of the integrand in Eq.~(\ref{vortexfinalcomx}) is a direct consequence of using the Laplacian as the simplest and most natural scalar differential operator in defining the coupling operators in Eq.~(\ref{couplefieldtheory}). It is quite remarkable that the detailed form of this integral leads to a vanishing $\lambda^*$ in three space dimensions and to the well-known leading-order coefficient in the $\epsilon$ expansion of $\lambda^*$. Equations (\ref{alpharesult}) and (\ref{wow}) moreover lead to an explicit result for the $\beta$ function (\ref{beta2nd}) which agrees with the famous result for the $\varphi^4$ theory as, for example, given in Eq.~(11.17) of \cite{ZinnJustin} or in Eq.~(18.5.7) of \cite{WeinbergQFT2}.

\section{Remarks on generalizations}
Our development of the quantum theory of a scalar field in $d$ space dimensions has been based on pairs of adjoint operators $a_{\bm{k}}^\dag$ and $a_{\bm{k}}$ creating and annihilating field quanta of momentum $\bm{k} \in \mathbb{R}^d$. We here consider generalizations based on on collections of adjoint operators ${a^J_{\bm{k}}}^\dag$ and $a^J_{\bm{k}}$ creating and annihilating field quanta of momentum $\bm{k}$, where $J$ is an additional index labeling the different types of quanta and the Fock space needs to be extended accordingly. For vector instead of scalar fields, as the simplest example, $J$ labels the different spatial components.

For bosonic fields, all creation operators commute among each other, and so do the annihilation operators. The only nontrivial commutation relations are
\begin{equation}\label{acommutatorgen}
    \Qcommu{a^J_{\bm{k}}}{{a^{J'}_{\bm{k}'}}^\dag} = \delta_{JJ'} \, \delta(\bm{k}-\bm{k}') .
\end{equation}

In terms of creation and annihilation operators, the Hamiltonian of the free field theory in the generalized setting is assumed to be of the form
\begin{equation}\label{H0gendef}
    H^{(0)} = \int \omega_{\bm{k}} \, {a^J_{\bm{k}}}^\dag a^J_{\bm{k}} \, d^dk ,
\end{equation}
where a summation over the index $J$ is implied by the Einstein summation convention and $\omega_{\bm{k}}$ is a non-negative real-valued function with the symmetry property $\omega_{-\bm{k}} = \omega_{\bm{k}}$. The standard form is given by the relativistic energy-momentum relationship (\ref{dispersion}), with a mass parameter $m$. For massless fields, we simply have $\omega_k = |\bm{k}|$. The fact that we assume $\omega_{\bm{k}}$ to be independent of $J$ (or that it depends in a restricted manner on $J$) expresses an underlying symmetry of the theory. In the same spirit, we can implement symmetry properties in the friction mechanism (however, Lorentz invariance would still be violated).

\subsection{Pure Yang-Mills fields}
For Yang-Mills fields, the label $J$ for the field quanta consists of a spatial index and an additional label associated with the infinitesimal generators of an underlying Lie group, where we use the letters $a,b,c$ for this kind of index [assuming $3$ values for $SU(2)$ corresponding to the W$^+$, W$^-$, and Z$^0$ bosons mediating weak interactions, $8$ values for $SU(3)$ corresponding to the gluon ``color octet'' mediating strong interactions, and $N^2-1$ values for general $SU(N)$]. More precisely, for Yang-Mills fields, the components are associated with the base vectors $T^a$ of the Lie algebra of the underlying linear Lie group (we here consider matrix groups only \cite{Rossmann}). We assume the orthonormality conditions
\begin{equation}\label{Liebase}
    2 \, {\rm tr} (T^a T^b) = \delta_{ab} .
\end{equation}
The most famous example are the three Pauli matrices as base vectors $T^a$ of the Lie algebra of $SU(2)$, which need to be divided by $2$ to satisfy Eq.~(\ref{Liebase}). The base vectors $T^a$ are traceless matrices.

In Yang-Mills theories, we deal with vector fields. The index $J$ hence is of the form $J=(\jmath \, a)$ where $\jmath$ is a spatial index and $a$ labels the base vectors of the Lie algebra. We use the amputated symbol $\jmath$ instead of $j$ to emphasize that this spatial index takes only $d-1$ instead of $d$ values. It is typical for gauge theories that one component of a vector can be eliminated by gauge transformations so that vectors are reduced to $d-1$ polarizations, here labeled by the amputated symbol $\jmath$. For the field quanta this implies that the vector bosons mediating weak and strong interactions in $d=3$ space dimensions come in two polarization states.

The spatial components of the physical fields and their conjugate momenta are obtained as generalizations of the Fourier transforms (\ref{phiexpression}) and (\ref{piexpression}),
\begin{equation}\label{gaugefieldsgen}
    A^{(j a)}(\bm{x}) = \frac{1}{\sqrt{2(2\pi)^d}} \int \frac{d^dk}{\sqrt{\omega_k}}
    \, \bm{e}^{(\jmath)}_{\bm{k} j} \left( {a^{(\jmath a)}_{\bm{k}}}^\dag
    + a^{(\jmath a)}_{-\bm{k}} \right) e^{- i \bm{k} \cdot \bm{x}} ,
\end{equation}
and
\begin{eqnarray}
    \Pi^{(j a)}(\bm{x}) &=& \frac{i}{\sqrt{2(2\pi)^d}} \int d^dk \sqrt{\omega_k}
    \left( \delta_{jl} - \frac{k_j k_l}{k^2} \right)
    \bm{e}^{(\jmath)}_{\bm{k} l} \nonumber \\
    & \times & \left( {a^{(\jmath a)}_{\bm{k}}}^\dag
    - a^{(\jmath a)}_{-\bm{k}} \right) e^{- i \bm{k} \cdot \bm{x}} ,
\label{gaugefieldsconj}
\end{eqnarray}
where summations over $\jmath, l$ are implied and the two polarization vectors $\bm{e}^{(1)}_{\bm{k}}$ and $\bm{e}^{(2)}_{\bm{k}}$ typically depend on $\bm{k}$, yet in a symmetric manner. The polarization vectors express certain gauge conditions. We here choose the particularly convenient axial gauge of a vanishing last field component, $A^{(d a)} = 0$, because it avoids some subtle complications known for other gauges (see remarks on p.~15 of \cite{WeinbergQFT2} and the added note in the conclusions of \cite{Mohapatra71a}). For the axial gauge in $d=3$ dimensions, we concretely choose \cite{Mohapatra71c}
\begin{equation}\label{axialgauge1}
    \bm{e}^{(1)}_{\bm{k}} = \frac{k}{k_3} \, \frac{1}{\sqrt{k_1^2+k_2^2}}
    \left( \begin{array}{c}
      k_1 \\
      k_2 \\
      0 \\
    \end{array} \right) ,
\end{equation}
and
\begin{equation}\label{axialgauge2}
    \bm{e}^{(2)}_{\bm{k}} = \frac{{\rm sgn}(k_3)}{\sqrt{k_1^2+k_2^2}}
    \left( \begin{array}{c}
      - k_2 \\
      k_1 \\
      0 \\
    \end{array} \right) .
\end{equation}
These vectors $\bm{e}^{(1)}_{\bm{k}}$ and $\bm{e}^{(2)}_{\bm{k}}$ are constructed such that one obtains canonical commutation relations for the fields $A^{(1 a)}$, $A^{(2 a)}$ and $\Pi^{(1 a)}$, $\Pi^{(2 a)}$,
\begin{equation}\label{transversecancom}
    \Qcommu{A^{(\imath a)}(\bm{x})}{\Pi^{(\jmath b)}(\bm{x}')} =
    i \, \delta_{\imath\jmath} \, \delta_{ab} \, \delta(\bm{x}-\bm{x}') .
\end{equation}
To obtain this generalization of Eq.~(\ref{acommutator}) for arbitrary $d$, we postulate the property
\begin{equation}\label{ekduald}
    \bm{e}^{(\jmath)}_{\bm{k} j}  \bm{e}^{(\jmath)}_{\bm{k} l} =
    \delta_{jl} + \frac{k_j k_l}{k_d^2} \quad
    \mbox{ for } j,l \in \{ 1, \ldots, d-1 \} ,
\end{equation}
for the polarization vectors of the axial gauge with $\bm{e}^{(\jmath)}_{\bm{k} d} = 0$, where summation over $\jmath$ but no integration over $\bm{k}$ is implied in Eq.~(\ref{ekduald}).

So far, we have made little use of the fact that we are dealing with a gauge theory: After fixing the gauge, $J=(\jmath \, a)$ has only $d-1$ spatial components labeled by $\jmath$, but a further label $a$ for the infinitesimal generator of the group occurs, and a so far unmotivated projector in momentum space has been introduced in Eq.~(\ref{gaugefieldsconj}). Much deeper use of the properties of gauge theories must, of course, be made in the formulation of the Hamiltonian for the interactions between the various kinds of field quanta. As a first step, we need to look at the following constraint arising for Yang-Mills fields,
\begin{equation}\label{algconstraint}
    \partial_j \Pi^{(j a)} + g f^{abc} A^{(j b)} \Pi^{(j c)} = 0 ,
\end{equation}
where $g$ is the coupling constant and $f^{abc}$ are the structure constants of the underlying Lie group, say $SU(n)$. For the free theory with $g=0$, the projector in Eq.~(\ref{gaugefieldsconj}) implies that the constraint (\ref{algconstraint}) is fulfilled so that its occurrence is now motivated. In the presence of interactions, we modify only the last component $\Pi^{(d a)}$ such that Eq.~(\ref{algconstraint}) is fulfilled. The components $\Pi^{(j a)}$ for $j \leq d-1$, which appear in the canonical commutation relations (\ref{transversecancom}), remain unchanged.

The constraint (\ref{algconstraint}), which fixes $\Pi^{(d a)}$, is a result of the field equations for the Yang-Mills theory. It depends on the strength of the interaction inherited from the Hamiltonian. This is very different from a gauge condition, say from fixing $A^{(d a)} = 0$. The gauge condition is an example of a primary constraint, whereas Eq.~(\ref{algconstraint}) is an example of a secondary constraint resulting from the consistency of the primary constraints with the evolution equations. Even more important is the distinction between first and second class constraints. Second class constraints can be treated in a consistent canonical quantization procedure based on the Dirac brackets of classical mechanics or field theory (see, for example, \cite{Dirac58a} or Sec.~7.6 of \cite{WeinbergQFT1}). First class constraints are typically related to unphysical degrees of freedom and can be taken into account by arbitrarily choosing a particular gauge. Unfortunately, there is no guarantee for the gauge invariance of the final results obtained by canonical quantization in different gauges. A loss of symmetry seems to be the unavoidable price to pay for a simple canonical quantization \cite{Dirac58b}.

Within the axial gauge, the full Hamiltonian can be expressed nicely in terms of the spatial components of the fields $A^{(j a)}$ and $\Pi^{(j a)}$ (see, for example,\ Eq.~(15.4.10) of \cite{WeinbergQFT2})
\begin{equation}\label{Hfulldef}
    H = \int \left[ \frac{1}{2} \Pi^{(j a)}(\bm{x}) \Pi^{(j a)}(\bm{x})
    + \frac{1}{4} F^{(ija)}(\bm{x}) F^{(ija)}(\bm{x}) \right] d^3x ,
\end{equation}
with the tensor fields
\begin{equation}\label{Ffieldsdef}
    F^{(ija)} = \partial^i A^{(j a)} - \partial^j A^{(i a)}
    + g f^{abc} A^{(i b)} A^{(j c)} .
\end{equation}
With the Fourier transforms (\ref{gaugefieldsgen}), (\ref{gaugefieldsconj}), the gauge condition $A^{(d a)} = 0$, and the modification of $\Pi^{(d a)}$ required to fulfill the constraint (\ref{algconstraint}), the Hamiltonian (\ref{Hfulldef}) can be rewritten in terms of the creation and annihilation operators ${a^J_{\bm{k}}}^\dag$ and $a^J_{\bm{k}}$. Except for an irrelevant constant contribution, the free Hamiltonian $H^{(0)}$ for $g=0$ is of the form (\ref{H0gendef}) provided that $\omega_k = |\bm{k}|$, that is, Yang-Mills fields must be massless. The interaction Hamiltonian contains first- and second-order terms in $g$ and a few field operators only, which is an extremely convenient feature of the axial gauge, in particular, for perturbation theory. Note that the form of the Hamiltonian describing the interactions depends on the gauge. It even happens that the interactions are described by a simple polynomial Hamiltonian in one gauge and by a nonpolynomial Hamiltonian in another gauge \cite{Mohapatra71a,Mohapatra71b,Mohapatra71c}.

With the described steps to introduce fields and Hamiltonian, it is possible to use the thermodynamic coarse-graining approach to the quantization of pure Yang-Mills fields. The introduction of friction simply happens by adding a label $J$ (to be summed over) to each of the creation and annihilation operators in Eq.~(\ref{QMEsym}). Also the coupling to matter would be straightforward. On the one hand, the loss of gauge invariance in the proposed procedure admittedly is a high price to pay. The results for gauge-invariant correlation functions, such as Wilson loops, need to be compared to conventional calculations and the $\beta$ function needs to be evaluated. On the other hand, a major benefit may be the simultaneous solution of all ultraviolet and infrared problems, where the latter are known to be particularly subtle for the present approach to Yang-Mills theories because axial gauges lead to additional singularities in the propagator so that naive principal-value recipes do not work \cite{CaraccioloCurMen82,LeroyMicheliRossi84,Landshoff86,LandshoffTaylor89,Leibbrandt90} and inconsistencies in the renormalization-group flow in perturbation theory might arise \cite{PanzaSoldati00,LitimPawlowski06}.

\subsection{Gravitation}
In the $\varphi^4$ and Yang-Mills theories, the friction mechanism provides a dynamic cutoff on the length scale $\ell$ introduced in Eq.~(\ref{smoothlength}). This smoothing length scale has no physical significance. We are dealing with an effective field theory valid only at length scales large compared to $\ell$. Physical predictions are obtained after removing the cutoff from the results of detailed calculations, that is, in the limit $\ell \rightarrow 0$. In the description of gravity, a quite different scenario should be expected. The smoothing length scale $\ell$ should be of the order of the Planck length and the dissipative friction mechanism should express a physical smearing of space and time at short scales, a fundamental effect that precludes further resolution. We hence expect an expression for the friction parameter $\gamma$ of the form
\begin{equation}\label{QGrel}
    \gamma \approx \frac{\hbar^2 G}{c \, \kB T_{\rm e}} ,
\end{equation}
where $G$ is the gravitational constant and a numerical prefactor still needs to be determined (we here show the occurrences of the reduced Planck constant and the speed of light explicitly). The determination of the prefactor could be based on a discussion of the entropy production rate as associated with black holes \cite{Bekenstein73,BardeenCarHawk73,Hawking74} or more general gravitational fields (for a thermodynamical discussion of the entropy of classical gravitational fields see \cite{hco180}). However, such wild speculations should be postponed until we will have succeeded in formulating a convincing covariant friction mechanism.

The above discussion implies that the role of renormalization-group theory is changed in quantum gravity. For the $\varphi^4$ and Yang-Mills theories, the $\beta$ function characterizes the running coupling constant of a minimal model that is appropriate to predict only the large-scale properties of its universality class. If the small regularizing length scale used for the minimal model goes to zero, we obtain the usual rules for handling the singularities of a field theory. In gravity, on the other hand, the small length scale $\ell$ has a physical meaning. We need a proper physical model on the scale $\ell$, not just a minimal model. No limit is required, no singularities occur. As gravity can be observed over an enormous range of length scales, the coupling constant at the Planck scale must be at its critical value. The Planck scale, the gravitational constant, and the friction parameter hence contain equivalent information, as is clear from the definition of the Planck scale and Eq.~(\ref{QGrel}). For gravity, renormalization is not related to fundamental aspects of constructing a field theory in terms of minimal models but belongs only to the realm of perturbation theory. When we pass form the Planck scale to larger scales, the original physical model is degraded to a minimal model for which, on the scales of the order of the scales of physical interest, we can apply perturbation theory with the hope to get useful results.

For the detailed mathematical formulation of the dissipative quantum theory of gravitation one might try to follow the procedure sketched for Yang-Mills theories. However, we are then faced with the same problems as in the canonical quantization of gravity. As a consequence of their complicated nonpolynomial functional form, the handling of the constraints resulting from gauge invariance is much more difficult for gravity. In particular, problems arise for the construction of the momentum representation. Ashtekar's famous idea of enlarging the gravitational phase space to incorporate spinors and to represent the spatial components of a metric in the decomposed form $g_{ij} = - {\sigma_{i A}}^B {\sigma_{j B}}^A$ (with summations over the $SU(2)$ spinor indices $A$ and $B$) in order to obtain simpler constraints \cite{Ashtekar86,Ashtekar87} (see, for example, also Sec.~3.2 of \cite{Nicolaietal05} and Sec.~4.3 of \cite{Kiefer}) may well be the key to obtain the dissipative quantum theory of gravitation.

\section{Summary and conclusions}
Within the framework provided by the thermodynamic quantum master equation, we propose to include irreversible terms into quantum field theory and to consider the limit of weak friction and low temperatures. These irreversible terms account for the fact that we need to eliminate degrees of freedom below certain length and time scales in order to be able to make use of the field idealization. The smoothing length scale resulting from the dissipative friction mechanism provides an alternative regularization scheme where an external cutoff is replaced by a dynamic process. However, the change from Hamiltonian to dissipative equations is much more than just another regularization scheme because it comes with a number of important implications. The vacuum state is no longer given by the ground state, but rather by a canonical density matrix characterized by a temperature $T_{\rm e}$. The Hamiltonian has a double role because it occurs both in the canonical density matrix and serves as a generator for the reversible contribution to dynamics. Entropy occurs naturally; it may be irrelevant in effective field theories because the entropy production rate is negligibly small, but it must play an important role in a full quantum theory of gravity.

In applying the proposed ideas to $\varphi^4$ theory in $d$ space dimensions, we have have elaborated explicitly that a perturbation theory can be constructed with guidance from a detailed-balance principle, without any need to go through Dyson's $U$ matrix to obtain meaningful results (see the discussion in Sec.~17.1 of \cite{BjorkenDrell}). All regularization is consistently provided by the irreversible contribution to time evolution that rapidly damps the local degrees of freedom. Although the irreversible friction mechanism is implemented in a non-covariant manner, the final physical predictions of second-order perturbation theory in the limit of vanishing friction coincide with the well-known manifestly covariant results. A covariant formulation of the friction mechanism would be desirable but, in view of the second-order derivatives occurring in Eq.~(\ref{couplefieldtheory}), it presumably requires the introduction of additional fields corresponding to spatial derivatives \cite{hcobet}.

In the present approach, renormalization-group theory is used as a tool to refine perturbation theory. It requires a certain characteristic occurrence of the small smoothing length scale in perturbation expansions, which we have verified explicitly to second order. As a result, we reproduce the well-known expression for the $\beta$ function for the running coupling constant. The fixed-point value of the dimensionless coupling constant is found to depend in a nontrivial way on space dimensionality.

Although we focus on perturbation expansions, the approach to quantum field theory proposed in the present paper is by no means restricted to perturbation theory. The fundamental quantum master equation (\ref{QMEsym}) could also be treated by non-perturbative methods, including stochastic simulation techniques. Not even the linearization of the master equation around equilibrium would be necessary if one liked to benefit from the full structure of the thermodynamic framework, for example, for fundamental or constructive developments. The underlying equation for the evolution of dissipative systems has a deep geometric structure and hence provides a sound basis for the description of dynamic quantum systems, including a Lyapunov function (entropy). The formulation is done according to the principles of nonequilibrium thermodynamics as the proper framework for systems with eliminated degrees of freedom.

For proving the new approach to be useful, one should establish its applicability to a wide range of quantum field theories. We have discussed how dissipative smoothing can be applied to pure Yang-Mills fields where we propose to do all calculations in the axial gauge. As a crucial test, the proposed dissipative friction mechanism should resolve the subtle infrared problems occurring for Yang-Mills fields in the axial gauge by providing the clear causality properties of irreversible equations. For the calculation of Wilson loops, the proper definition of multi-time correlations needs to be understood, including the proper implementation of time-reversal symmetry, detailed balance, and their preservation under perturbation theory. For quantum gravity, the frictional smoothing mechanism would need to be elevated from a regularization scheme to a physical smearing of space and time.

As a first step, we have restricted ourselves to second-order perturbation expansions in $\varphi^4$ theory. Of course, the dissipative approach to quantum field theory should also be validated for higher orders of perturbation theory and for more complicated field theories. One might want to attempt to establish a structure of perturbation expansions consistent with renormalization-group theory by analyzing Feynman diagrams. However, in view of the robust geometric structure of the thermodynamic approach, a deeper discussion of the limits of vanishing friction and temperature seems to be the more relevant issue. To simplify the practical calculation of perturbation expansions, one needs to formulate rules to identify those Feynman diagrams that do not contribute in the limits of vanishing friction and temperature. Many details of dissipative quantum field theory still need to be clarified, many possibilities still need to be explored.

\begin{acknowledgments}
I would like to thank J\"urg Fr\"ohlich and Gian Michele Graf for very critical, thought-provoking comments on this work.
\end{acknowledgments}

\appendix

\section{Some useful integrals} \label{app_integrals}
We here collect a number of useful integrals that occur in the perturbation expansion of $\varphi^4$ theory. To facilitate a direct comparison, we introduce the integrals $I_1$, $I_2$, $I_3$ in exactly the same way as in Sec.~11.5.1 of \cite{ZinnJustin}. These integrals depend only on the dimensionality $d$. As before, we use $d$ for the dimension of space, $D=d+1$ for the dimension of spacetime, and $\epsilon = 3-d = 4-D$. The $D=d+1$ components of a spacetime vector $\underline{k}$ are given by $\underline{k} = (\kappa, \bm{k})$, where $\kappa$ is real and $\bm{k}$ has $d$ components. Connecting integrals in $d$ and $D$ dimensions, where we consider both real and imaginary time, is the key to revealing Lorentz invariance.

The simplest integral is defined by (see Eq.~(11.39) of \cite{ZinnJustin})
\begin{eqnarray}
    I_1 &=& \frac{m^{1-d}}{(2\pi)^d} \, \int \frac{d^dk}{2 \omega_k}
    \nonumber\\
    &=& \frac{m^{2-D}}{(2\pi)^D} \, \int d^Dk \frac{i}{\kappa^2-\omega_k^2+i\varepsilon}
    \nonumber\\
    &=& \frac{m^{2-D}}{(2\pi)^D} \, \int d^Dk \frac{1}{\kappa^2+\omega_k^2} .
\label{IntegralI1}
\end{eqnarray}
Note that the integral $I_1$ may be ill-defined in certain integer dimensions. If $\kappa$ corresponds to the real frequency, one needs to be careful with the proper handling of the poles which are located on the integration line; then, the $i\varepsilon$ with infinitesimal positive $\varepsilon$ shifts the poles to avoid ambiguities. If $\kappa$ corresponds to the imaginary part of the frequency, the integration path is well separated from the poles and the $i\varepsilon$ prescription becomes irrelevant. After properly closing the contour for the $\kappa$ integration, Cauchy's integral theorem of complex analysis allows us to pass from the $D$-dimensional integrals to the $d$-dimensional representation.

By means of the general formula for reducing $d$-dimensional spherically symmetric integrals to one-dimensional integrals,
\begin{equation}\label{dintto1int}
    \int f(k) \, d^dk = \frac{2 \pi^{d/2}}{\Gamma(d/2)}  \int_0^\infty f(k) \, k^{d-1} dk ,
\end{equation}
we obtain the explicit result
\begin{equation}\label{IntegralI1expl}
    I_1 = - \frac{1}{(2\pi)^2} \frac{(2\sqrt{\pi})^\epsilon}{\epsilon(2-\epsilon)} \,
    \Gamma\left( 1 + \frac{\epsilon}{2} \right) .
\end{equation}
The following two useful integrals are closely related to $I_1$,
\begin{equation}\label{IntegralI1a}
    I'_1  = \frac{m^{3-d}}{(2\pi)^d} \, \int \frac{d^dk}{2 \omega_k^3} =
    \frac{1}{(2\pi)^2} \frac{(2\sqrt{\pi})^\epsilon}{\epsilon} \,
    \Gamma\left( 1 + \frac{\epsilon}{2} \right) ,
\end{equation}
and
\begin{eqnarray}
    I''_1  &=& \frac{m^{3-d}}{(2\pi)^d} \, \int \frac{d^dk}{2 \omega_k^2 (\omega_k + m)} =
    \frac{1}{(2\pi)^2} \frac{(2\sqrt{\pi})^\epsilon}{\epsilon(1-\epsilon)} \nonumber\\
    &\times& \left[ \Gamma\left( 1 + \frac{\epsilon}{2} \right)
    - \frac{\epsilon}{2} \, \sqrt{\pi} \, \Gamma\left( \frac{1}{2} + \frac{\epsilon}{2} \right)
    \right] .
\label{IntegralI1b}
\end{eqnarray}

As a next step, we consider the dimensionless integral
\begin{eqnarray}
    I_{\bm{k}}(-\kappa^2) &=&  \frac{m^{6-2D}}{(2\pi)^{2D}} \int d^Dk_1 d^Dk_2 d^Dk_3
    \nonumber\\
    &\times& \frac{\delta(\underline{k}_1+\underline{k}_2+\underline{k}_3
    +\underline{k})}{(\kappa_1^2+\omega_{k_1}^2)
    (\kappa_2^2+\omega_{k_2}^2)(\kappa_3^2+\omega_{k_3}^2)} , \qquad
\label{Integralgenexp3}
\end{eqnarray}
where, again, certain integer values of $D$ should be excluded. This integral has been discussed in great detail on pp.~282--284 of \cite{ZinnJustin}. According to Eqs.~(11.38), (11.40), and (11.41) of \cite{ZinnJustin}, the useful integrals $I_2$ and $I_3$ can be introduced by expanding $I_{\bm{k}}(\omega^2)$ in terms of $\omega^2$,
\begin{equation}\label{IntegralI2I3}
    I_{\bm{k}}(\omega^2) \Big|_{\bm{k}=0} =
   I_2 -I_3 \, \frac{\omega^2}{m^2} + O(\omega^4) .
\end{equation}
By deforming the integration path from imaginary to real frequencies and carefully stating by the $i\varepsilon$ prescription how poles on the new integration line have to be circumvented, we arrive at the Lorentz invariant expression occurring, for example, in truncated form for $D=4$ on p.~174 of \cite{Zee},
\begin{eqnarray}
    I_{\bm{k}}(\kappa^2) &=&  \frac{m^{6-2D}}{(2\pi)^{2D}} \int d^Dk_1 d^Dk_2 d^Dk_3
    \nonumber\\
    &\times& \!\! \frac{\delta(\underline{k}_1+\underline{k}_2+\underline{k}_3
    +\underline{k})}{(\kappa_1^2-\omega_{k_1}^2+i\varepsilon)
    (\kappa_2^2-\omega_{k_2}^2+i\varepsilon)(\kappa_3^2-\omega_{k_3}^2+i\varepsilon)} .
    \nonumber\\ &&
\label{Integralgenexp2}
\end{eqnarray}
Again, we can pass from $D$-dimensional to $d$-dimensional integrals. After replacing the factor $\delta(\kappa_1+\kappa_2+\kappa_3 +\kappa)$ contained in Eq.~(\ref{Integralgenexp2}) by its one-dimensional Fourier transform, the integrations over $\kappa_1$, $\kappa_2$, and $\kappa_3$ can be performed. After a further integration of an exponential, we obtain
\begin{eqnarray}
    I_{\bm{k}}(\omega^2) &=&  \frac{m^{4-2d}}{(2\pi)^{2d}}
    \int \frac{d^dk_1 d^dk_2 d^dk_3}{4 \, \omega_{k_1} \omega_{k_2} \omega_{k_3}} \,
    \delta(\bm{k}_1+\bm{k}_2+\bm{k}_3+\bm{k})
    \nonumber\\ &\times&
    \frac{\omega_{k_1}+\omega_{k_2}+\omega_{k_3}}{
    (\omega_{k_1}+\omega_{k_2}+\omega_{k_3}-i\varepsilon)^2 - \omega^2} .
\label{Integralgenexp1}
\end{eqnarray}

For the actual evaluation of $I_{\bm{k}}(\omega^2)$,  the form (\ref{Integralgenexp3}) offers the most convenient starting point. Note that each of the factors in the denominator of Eq.~(\ref{Integralgenexp3}) contains a positive-definite quadratic form $Q$ of the integration variables. After using the identity $1/Q = \int_0^\infty e^{-Qz} dz$ for each of the three factors in the denominator, the two Gaussian $D$-dimensional integrations remaining after making use of the $\delta$ function can be carried out easily and only three one-dimensional integrations remain to be done (see Sec.~9.6 of \cite{ZinnJustin}).

In comparing our results to those of \cite{ZinnJustin}, for example, the formula for $I_1$ given in Eq.~(\ref{IntegralI1expl}) to (11.44) of \cite{ZinnJustin}, the following identities are useful (see 8.334 of \cite{GradshteynRyzhik}):
\begin{equation}\label{GamGamsin}
    \Gamma(1-x) \, \Gamma(x) = \frac{\pi}{\sin \pi x} ,
\end{equation}
\begin{equation}\label{GamGamcos}
    \Gamma\left(\frac{1}{2}-x\right) \, \Gamma\left(\frac{1}{2}+x\right) =
    \frac{\pi}{\cos \pi x} .
\end{equation}

\section{Some useful examples of $\Gamma_A$} \label{app_GammaAexamples}
Among the normal ordered products of up to three creation/annihilation operators, $\Gamma_A$ vanishes for all operators except for $a^\dag_{\bm{k}_1} a_{\bm{k}_2}$, $a^\dag_{\bm{k}_1} a_{\bm{k}_2} a_{\bm{k}_3}$, and $a^\dag_{\bm{k}_1} a^\dag_{\bm{k}_2} a_{\bm{k}_3}$. For the product of two operators, we have
\begin{equation}\label{GammaAexpl2}
  \Gamma_{a^\dag_{\bm{k}_1} a_{\bm{k}_2}}
  = w(\omega_{k_1}) \, \frac{2 \gamma_{k_1}}{\beta \omega_{k_1}} \,
    \delta(\bm{k}_1-\bm{k}_2) .
\end{equation}
For the product of three creation/annihilation operators, we find
\begin{eqnarray}
    \Gamma_{a^\dag_{\bm{k}_1} a_{\bm{k}_2} a_{\bm{k}_3}} &=&
    \frac{\gamma_{k_1}}{e^{\beta\omega_{k_1}} - 1}
    \nonumber\\
    &\times&  \Big\{ \delta(\bm{k}_1-\bm{k}_2) \, \Big[ W(\omega_{k_1},-\omega_{k_1}-\omega_{k_3})
    \nonumber\\
    && + \, W(-\omega_{k_1},\omega_{k_1}-\omega_{k_3}) \Big] a_{\bm{k}_3}
    \nonumber\\
    && + \, \delta(\bm{k}_1-\bm{k}_3) \, \Big[ W(\omega_{k_1},-\omega_{k_1}-\omega_{k_2})
    \nonumber\\
    && + \, W(-\omega_{k_1},\omega_{k_1}-\omega_{k_2}) \Big] a_{\bm{k}_2} \Big\} ,
\label{GammaAexpl3a}
\end{eqnarray}
and
\begin{eqnarray}
    \Gamma_{a^\dag_{\bm{k}_1} a^\dag_{\bm{k}_2} a_{\bm{k}_3}} &=&
    \frac{\gamma_{k_3}}{e^{\beta\omega_{k_3}} - 1}
    \nonumber\\
    &\times&  \Big\{ \delta(\bm{k}_2-\bm{k}_3) \, \Big[ W(\omega_{k_3},\omega_{k_1}-\omega_{k_3})
    \nonumber\\
    && + \, W(-\omega_{k_3},\omega_{k_1}+\omega_{k_3}) \Big] a^\dag_{\bm{k}_1}
    \nonumber\\
    && + \, \delta(\bm{k}_1-\bm{k}_3) \, \Big[ W(\omega_{k_3},\omega_{k_2}-\omega_{k_3})
    \nonumber\\
    && + \, W(-\omega_{k_3},\omega_{k_2}+\omega_{k_3}) \Big] a^\dag_{\bm{k}_2} \Big\} .
\label{GammaAexpl3b}
\end{eqnarray}
These results are obtained by straightforward evaluation of Eq.~(\ref{GammaAexpl}).

\begin{widetext}

\section{Some useful commutators}
To facilitate the calculation of averages occurring in perturbation theory, we calculate some basic commutators involving the Hamiltonian. They are obtained in a straightforward manner from the fundamental commutation relations (\ref{acommutator}) and the from (\ref{H1defNO}) of the $\varphi^4$ Hamiltonian. The following single commutators are used in evaluating the averages in Eqs.~(\ref{pertYkdef}) and (\ref{Gammaperturbation2}):
\begin{eqnarray}
    \Qcommu{a_{\bm{k}}}{H^{(1)}} &=& \frac{\lambda z}{\tilde{\omega}_k}
    \Big( a_{\bm{k}} + a^\dag_{-\bm{k}} \Big)
    + \frac{\lambda}{24} \frac{1}{(2\pi)^d}
    \frac{1}{\sqrt{\tilde{\omega}_k}}
    \int \frac{d^dk_1 d^dk_2 d^dk_3}{\sqrt{\tilde{\omega}_{k_1}
    \tilde{\omega}_{k_2} \tilde{\omega}_{k_3}}} \,
    \delta(\bm{k}_1+\bm{k}_2+\bm{k}_3+\bm{k}) \nonumber \\
    && \times  \Big( a_{-\bm{k}_1} a_{-\bm{k}_2} a_{-\bm{k}_3}
    + 3 a^\dag_{\bm{k}_1} a_{-\bm{k}_2} a_{-\bm{k}_3}
    + 3 a^\dag_{\bm{k}_1} a^\dag_{\bm{k}_2} a_{-\bm{k}_3}
    + a^\dag_{\bm{k}_1} a^\dag_{\bm{k}_2} a^\dag_{\bm{k}_3}
    \Big) ,
\label{H1_commu_a}
\end{eqnarray}
and
\begin{eqnarray}
    \Qcommu{H^{(1)}}{a^\dag_{\bm{k}}} &=& \frac{\lambda z}{\tilde{\omega}_k}
    \Big( a^\dag_{\bm{k}} + a_{-\bm{k}} \Big)
    + \frac{\lambda}{24} \frac{1}{(2\pi)^d}
    \frac{1}{\sqrt{\tilde{\omega}_{k}}}
    \int \frac{d^dk_1 d^dk_2 d^dk_3}{\sqrt{\tilde{\omega}_{k_1}
    \tilde{\omega}_{k_2} \tilde{\omega}_{k_3}}} \,
    \delta(\bm{k}_1+\bm{k}_2+\bm{k}_3-\bm{k}) \nonumber \\
    && \times  \Big( a_{-\bm{k}_1} a_{-\bm{k}_2} a_{-\bm{k}_3}
    + 3 a^\dag_{\bm{k}_1} a_{-\bm{k}_2} a_{-\bm{k}_3}
    + 3 a^\dag_{\bm{k}_1} a^\dag_{\bm{k}_2} a_{-\bm{k}_3}
    + a^\dag_{\bm{k}_1} a^\dag_{\bm{k}_2} a^\dag_{\bm{k}_3}
    \Big) .
\label{H1_commu_ad}
\end{eqnarray}
The following type of identities for single commutators allows us the separation of the two contributions represented by the Feynman diagrams in Fig.~\ref{fig_feynman_diagr4}(a) and (b):
\begin{equation}\label{H1_commu_aa}
    \Qcommu{a_{\bm{k}_1} a_{\bm{k}_2}}{H^{(1)}} =
    \Qcommu{a_{\bm{k}_1}}{H^{(1)}} \, a_{\bm{k}_2} +
    \Qcommu{a_{\bm{k}_2}}{H^{(1)}} \, a_{\bm{k}_1} +
    \Qcommu{a_{\bm{k}_1}}{\Qcommu{a_{\bm{k}_2}}{H^{(1)}}} ,
\end{equation}
and
\begin{equation}\label{H1_commu_adad}
    \Qcommu{H^{(1)}}{a^\dag_{\bm{k}_1} a^\dag_{\bm{k}_2}} =
    a^\dag_{\bm{k}_1} \, \Qcommu{H^{(1)}}{a^\dag_{\bm{k}_2}} +
    a^\dag_{\bm{k}_2} \, \Qcommu{H^{(1)}}{a^\dag_{\bm{k}_1}} +
    \Qcommu{\Qcommu{H^{(1)}}{a^\dag_{\bm{k}_1}}}{a^\dag_{\bm{k}_2}} .
\end{equation}

We also provide a number of double commutators. The following double commutators are used in Eqs.~(\ref{pertXkdef}) and (\ref{Gammaperturbation1}):
\begin{eqnarray}
    \Qcommu{a_{\bm{k}_1}}{\Qcommu{a_{\bm{k}_2}}{H^{(1)}}} &=&
    \frac{\lambda}{8} \frac{1}{(2\pi)^d}
    \frac{1}{\sqrt{\tilde{\omega}_{k_1} \tilde{\omega}_{k_2}}}
    \int \frac{d^dk'_1 d^dk'_2}{\sqrt{\tilde{\omega}_{k'_1} \tilde{\omega}_{k'_2}}} \,
    \delta(\bm{k}'_1+\bm{k}'_2+\bm{k}_1+\bm{k}_2) \,
    \Big( a_{-\bm{k}'_1} a_{-\bm{k}'_2}
    + 2 a^\dag_{\bm{k}'_1} a_{-\bm{k}'_2}
    + a^\dag_{\bm{k}'_1} a^\dag_{\bm{k}'_2} \Big) \nonumber\\
    &+& \frac{\lambda z}{\tilde{\omega}_{k_1}} \, \delta(\bm{k}_1+\bm{k}_2) ,
\label{H1_dcommu_aa}
\end{eqnarray}
\begin{eqnarray}
    \Qcommu{\Qcommu{H^{(1)}}{a^\dag_{\bm{k}_1}}}{a^\dag_{\bm{k}_2}} &=&
    \frac{\lambda}{8} \frac{1}{(2\pi)^d}
    \frac{1}{\sqrt{\tilde{\omega}_{k_1} \tilde{\omega}_{k_2}}}
    \int \frac{d^dk'_1 d^dk'_2}{\sqrt{\tilde{\omega}_{k'_1} \tilde{\omega}_{k'_2}}} \,
    \delta(\bm{k}'_1+\bm{k}'_2-\bm{k}_1-\bm{k}_2) \,
    \Big( a_{-\bm{k}'_1} a_{-\bm{k}'_2}
    + 2 a^\dag_{\bm{k}'_1} a_{-\bm{k}'_2}
    + a^\dag_{\bm{k}'_1} a^\dag_{\bm{k}'_2} \Big) \nonumber\\
    &+& \frac{\lambda z}{\tilde{\omega}_{k_1}} \, \delta(\bm{k}_1+\bm{k}_2) ,
\label{H1_dcommu_adad}
\end{eqnarray}
and
\begin{eqnarray}
    \Qcommu{a_{\bm{k}_1}}{\Qcommu{H^{(1)}}{a^\dag_{\bm{k}_2}}} &=&
    \frac{\lambda}{8} \frac{1}{(2\pi)^d}
    \frac{1}{\sqrt{\tilde{\omega}_{k_1} \tilde{\omega}_{k_2}}}
    \int \frac{d^dk'_1 d^dk'_2}{\sqrt{\tilde{\omega}_{k'_1} \tilde{\omega}_{k'_2}}} \,
    \delta(\bm{k}'_1+\bm{k}'_2+\bm{k}_1-\bm{k}_2) \,
    \Big( a_{-\bm{k}'_1} a_{-\bm{k}'_2}
    + 2 a^\dag_{\bm{k}'_1} a_{-\bm{k}'_2}
    + a^\dag_{\bm{k}'_1} a^\dag_{\bm{k}'_2} \Big) \nonumber\\
    &+& \frac{\lambda z}{\tilde{\omega}_{k_1}} \, \delta(\bm{k}_1-\bm{k}_2) .
\label{H1_dcommu_aad}
\end{eqnarray}
The following type of identities for double commutators facilitates the evaluation of the average in Eq.~(\ref{perturbationfoc2a}):
\begin{eqnarray}
    \Qcommu{a_{\bm{k}_1} a_{\bm{k}_2}}{\Qcommu{H^{(1)}}{a^\dag_{\bm{k}'_1} a^\dag_{\bm{k}'_2}}} &=&
    \nonumber \\ && \hspace{-10em}
    \delta(\bm{k}_1-\bm{k}'_1) \, a_{\bm{k}_2} \Qcommu{H^{(1)}}{a^\dag_{\bm{k}'_2}}
    + \delta(\bm{k}_1-\bm{k}'_2) \, a_{\bm{k}_2} \Qcommu{H^{(1)}}{a^\dag_{\bm{k}'_1}}
    + \delta(\bm{k}_2-\bm{k}'_1) \, a_{\bm{k}_1} \Qcommu{H^{(1)}}{a^\dag_{\bm{k}'_2}}
    + \delta(\bm{k}_2-\bm{k}'_2) \, a_{\bm{k}_1} \Qcommu{H^{(1)}}{a^\dag_{\bm{k}'_1}}
    \nonumber \\ && \hspace{-10em}
    + a^\dag_{\bm{k}'_1} \, \Qcommu{a_{\bm{k}_2}}{\Qcommu{H^{(1)}}{a^\dag_{\bm{k}'_2}}} \, a_{\bm{k}_1}
    + a^\dag_{\bm{k}'_2} \, \Qcommu{a_{\bm{k}_2}}{\Qcommu{H^{(1)}}{a^\dag_{\bm{k}'_1}}} \, a_{\bm{k}_1}
    + a^\dag_{\bm{k}'_1} \, \Qcommu{a_{\bm{k}_1}}{\Qcommu{H^{(1)}}{a^\dag_{\bm{k}'_2}}} \, a_{\bm{k}_2}
    + a^\dag_{\bm{k}'_2} \, \Qcommu{a_{\bm{k}_1}}{\Qcommu{H^{(1)}}{a^\dag_{\bm{k}'_1}}} \, a_{\bm{k}_2}
    \nonumber \\ && \hspace{-10em}
    + a^\dag_{\bm{k}'_1} \, \Qcommu{a_{\bm{k}_1}}{\Qcommu{a_{\bm{k}_2}}{\Qcommu{H^{(1)}}{a^\dag_{\bm{k}'_2}}}}
    + a^\dag_{\bm{k}'_2} \, \Qcommu{a_{\bm{k}_1}}{\Qcommu{a_{\bm{k}_2}}{\Qcommu{H^{(1)}}{a^\dag_{\bm{k}'_1}}}}
    + \Qcommu{a_{\bm{k}_1}}{\Qcommu{\Qcommu{H^{(1)}}{a^\dag_{\bm{k}'_1}}}{a^\dag_{\bm{k}'_2}}} \, a_{\bm{k}_2}
    + \Qcommu{a_{\bm{k}_2}}{\Qcommu{\Qcommu{H^{(1)}}{a^\dag_{\bm{k}'_1}}}{a^\dag_{\bm{k}'_2}}} \, a_{\bm{k}_1}
    \nonumber \\ && \hspace{-10em}
    + \Qcommu{a_{\bm{k}_1}}{\Qcommu{a_{\bm{k}_2}}{\Qcommu{\Qcommu{H^{(1)}
    }{a^\dag_{\bm{k}'_1}}}{a^\dag_{\bm{k}'_2}}}} ,
\label{H1_dcommu_aaadad}
\end{eqnarray}
and
\begin{eqnarray}
    \Qcommu{\Qcommu{a_{\bm{k}_1} a_{\bm{k}_2}}{H^{(1)}}}{a^\dag_{\bm{k}'_1} a^\dag_{\bm{k}'_2}} &=&
    \nonumber \\ && \hspace{-10em}
    \delta(\bm{k}_1-\bm{k}'_1) \, \Qcommu{a_{\bm{k}_2}}{H^{(1)}} a^\dag_{\bm{k}'_2}
    + \delta(\bm{k}_1-\bm{k}'_2) \, \Qcommu{a_{\bm{k}_2}}{H^{(1)}} a^\dag_{\bm{k}'_1}
    + \delta(\bm{k}_2-\bm{k}'_1) \, \Qcommu{a_{\bm{k}_1}}{H^{(1)}} a^\dag_{\bm{k}'_2}
    + \delta(\bm{k}_2-\bm{k}'_2) \, \Qcommu{a_{\bm{k}_1}}{H^{(1)}} a^\dag_{\bm{k}'_1}
    \nonumber \\ && \hspace{-10em}
    + a^\dag_{\bm{k}'_1} \, \Qcommu{a_{\bm{k}_2}}{\Qcommu{H^{(1)}}{a^\dag_{\bm{k}'_2}}} \, a_{\bm{k}_1}
    + a^\dag_{\bm{k}'_2} \, \Qcommu{a_{\bm{k}_2}}{\Qcommu{H^{(1)}}{a^\dag_{\bm{k}'_1}}} \, a_{\bm{k}_1}
    + a^\dag_{\bm{k}'_1} \, \Qcommu{a_{\bm{k}_1}}{\Qcommu{H^{(1)}}{a^\dag_{\bm{k}'_2}}} \, a_{\bm{k}_2}
    + a^\dag_{\bm{k}'_2} \, \Qcommu{a_{\bm{k}_1}}{\Qcommu{H^{(1)}}{a^\dag_{\bm{k}'_1}}} \, a_{\bm{k}_2}
    \nonumber \\ && \hspace{-10em}
    + a^\dag_{\bm{k}'_1} \, \Qcommu{a_{\bm{k}_1}}{\Qcommu{a_{\bm{k}_2}}{\Qcommu{H^{(1)}}{a^\dag_{\bm{k}'_2}}}}
    + a^\dag_{\bm{k}'_2} \, \Qcommu{a_{\bm{k}_1}}{\Qcommu{a_{\bm{k}_2}}{\Qcommu{H^{(1)}}{a^\dag_{\bm{k}'_1}}}}
    + \Qcommu{a_{\bm{k}_1}}{\Qcommu{\Qcommu{H^{(1)}}{a^\dag_{\bm{k}'_1}}}{a^\dag_{\bm{k}'_2}}} \, a_{\bm{k}_2}
    + \Qcommu{a_{\bm{k}_2}}{\Qcommu{\Qcommu{H^{(1)}}{a^\dag_{\bm{k}'_1}}}{a^\dag_{\bm{k}'_2}}} \, a_{\bm{k}_1}
    \nonumber \\ && \hspace{-10em}
    + \Qcommu{a_{\bm{k}_1}}{\Qcommu{a_{\bm{k}_2}}{\Qcommu{\Qcommu{H^{(1)}
    }{a^\dag_{\bm{k}'_1}}}{a^\dag_{\bm{k}'_2}}}} .
\label{H1_dcommu_aaadadx}
\end{eqnarray}

\section{Exact second-order perturbation results for propagator}\label{app_exact2ndord}
By evaluating the average (\ref{pertXkdef}) by means of the first order expansion (\ref{averageperturb}) of averages, the double commutator (\ref{H1_dcommu_aad}), and Wick's theorem (\ref{Wick}), we obtain the exact result
\begin{equation}\label{X2full}
    X_{\bm{k}} = \lambda \, \frac{J+2z}{2\tilde{\omega}_k}
    - \lambda^2 \, \frac{\beta J}{16 \tilde{\omega}_k}
    \frac{1}{(2\pi)^d} \int \frac{d^dq}{\tilde{\omega}_q^2}
    \left[ \frac{w(2\omega_q)}{(1-e^{-\beta\omega_q})^2}
    + \frac{2}{(e^{\beta\omega_q}-1)(1-e^{-\beta\omega_q})}
    + \frac{w(-2\omega_q)}{(e^{\beta\omega_q}-1)^2} \right] ,
\end{equation}
with
\begin{equation}\label{Jdddef}
    J = \frac{1}{(2\pi)^d} \int \frac{d^dq}{2\tilde{\omega}_q (e^{\beta\omega_q}-1)} .
\end{equation}
Note that, in the low-temperature limit $\beta \rightarrow \infty$, the integral $J$ becomes exponentially small. In a similar way, the average (\ref{pertYkdef}) can be evaluated by means of the commutators (\ref{H1_commu_a}) and (\ref{H1_commu_ad}), the formula (\ref{R0barreduction}) with the explicit expressions of Appendix~\ref{app_GammaAexamples}, and Wick's theorem (\ref{Wick}). The complete result is
\begin{eqnarray}
    Y_{\bm{k}}(\omega) &=&
    - \lambda^2 \, \frac{z(J+2z)}{2 \, \tilde{\omega}_k^2}
    \bigg( \frac{1}{\omega+\omega_k-i\gamma_k} - \frac{1}{\omega-\omega_k-i\gamma_k} \bigg)
    \nonumber\\
    &-& \lambda^2 \, \frac{J+2z}{8 \, \tilde{\omega}_k^2}
    \frac{1}{(2\pi)^d} \int \frac{d^dq}{\tilde{\omega}_q (e^{\beta\omega_q}-1)}
    \bigg( \frac{1}{\omega+\omega_k-i\gamma_{a^\dag_{\bm{q}} a_{\bm{q}} a_{\bm{k}}}}
    - \frac{1}{\omega-\omega_k-i\gamma_{a^\dag_{-\bm{k}} a^\dag_{\bm{q}} a_{\bm{q}}}}
    \bigg)
    \nonumber\\
    &+& \lambda^2 \, \frac{J+2z}{8 \, \tilde{\omega}_k^2}
    \frac{1}{(2\pi)^d} \int \frac{i \gamma_q \, \big[ W(\omega_q,-\omega_q+\omega_k)
    + W(\omega_q,-\omega_q-\omega_k) \big] \, d^dq}{\tilde{\omega}_q
    (e^{\beta\omega_q}-1)}
    \bigg( \frac{1}{\omega+\omega_k-i\gamma_{a^\dag_{\bm{q}} a_{\bm{q}} a_{\bm{k}}}}
    \, \frac{1}{\omega+\omega_k-i\gamma_k}
    \nonumber\\
    && \hspace{8em} - \,
    \frac{1}{\omega-\omega_k-i\gamma_{a^\dag_{-\bm{k}} a^\dag_{\bm{q}} a_{\bm{q}}}}
    \, \frac{1}{\omega-\omega_k-i\gamma_k}
    \bigg)
    \nonumber\\
    &-& \lambda^2 \, \frac{1}{96 \, \tilde{\omega}_k}
    \frac{1}{(2\pi)^{2d}} \int \frac{d^dk_1 d^dk_2 d^dk_3}{\tilde{\omega}_{k_1}
    \tilde{\omega}_{k_2} \tilde{\omega}_{k_3}} \,
    \frac{\delta(\bm{k}_1+\bm{k}_2+\bm{k}_3+\bm{k})}{(1-e^{-\beta\omega_{k_1}})
    (1-e^{-\beta\omega_{k_2}})(1-e^{-\beta\omega_{k_3}})}
    \nonumber\\
    &\times& \bigg[
    \frac{1-e^{-\beta(\omega_{k_1}+\omega_{k_2}+\omega_{k_3})}}{
    \omega+\omega_{k_1}+\omega_{k_2}+\omega_{k_3}
    -i\gamma_{a_{-\bm{k}_1} a_{-\bm{k}_2} a_{-\bm{k}_3}}}
    - \frac{1-e^{-\beta(\omega_{k_1}+\omega_{k_2}+\omega_{k_3})}}{
    \omega-\omega_{k_1}-\omega_{k_2}-\omega_{k_3}
    -i\gamma_{a^\dag_{\bm{k}_1} a^\dag_{\bm{k}_2} a^\dag_{\bm{k}_3}}}
    \nonumber\\
    &+& 3 \, \frac{e^{-\beta\omega_{k_1}}-e^{-\beta(\omega_{k_2}+\omega_{k_3})}}{
    \omega-\omega_{k_1}+\omega_{k_2}+\omega_{k_3}
    -i\gamma_{a^\dag_{\bm{k}_1} a_{-\bm{k}_2} a_{-\bm{k}_3}}}
    -  3 \, \frac{e^{-\beta\omega_{k_3}}-e^{-\beta(\omega_{k_1}+\omega_{k_2})}}{
    \omega-\omega_{k_1}-\omega_{k_2}+\omega_{k_3}
    -i\gamma_{a^\dag_{\bm{k}_1} a^\dag_{\bm{k}_2} a_{-\bm{k}_3}}} \bigg] .
\label{Y2full}
\end{eqnarray}

\end{widetext}



\end{document}